\documentclass[superscriptaddress, amsmath,amssymb,
 aps, jcp]{revtex4}

\usepackage{graphicx}% Include figure files
\usepackage{dcolumn}% Align table columns on decimal point

\usepackage{bm}% bold math
\usepackage[english]{babel}

\usepackage[utf8]{inputenc}
\usepackage[T1]{fontenc}
\usepackage{mathtools}
\usepackage{placeins}

\usepackage{lipsum}

\usepackage{letltxmacro}

\LetLtxMacro{\ORIGselectlanguage}{\selectlanguage}
\makeatletter
\DeclareRobustCommand{\selectlanguage}[1]{%
  \@ifundefined{alias@\string#1}
    {\ORIGselectlanguage{#1}}
    {\begingroup\edef\x{\endgroup
       \noexpand\ORIGselectlanguage{\@nameuse{alias@#1}}}\x}%
}
\newcommand{\definelanguagealias}[2]{%
  \@namedef{alias@#1}{#2}%
}
\makeatother

\definelanguagealias{en}{english}

\newcommand{\rref}[1]{Eq.\ (\ref{#1})}
\newcommand{\rrefsa}[1]{Eqs.\ (\ref{#1})}
\newcommand{\rrefsb}[1]{(\ref{#1})}
\newcommand{\bos}[1]{\boldsymbol{#1}}
\newcommand{\bR}{\bos{R}}
\newcommand{\bA}{\bos{A}}
\newcommand{\ubA}{\underline{\bos{A}}}

\newcommand{\bp}{\bos{p}}

\newcommand{\bs}{\bos{s}}

\newcommand{\br}{\bos{r}}

\newcommand{\be}{\bos{e}}

\newcommand{\bJ}{\bos{J}}

\usepackage{dcolumn}%
\newcolumntype{.}{D{.}{.}{8}}

\DeclareUnicodeCharacter{2009}{\,}
\usepackage{bm}
\usepackage{mathtools}
\usepackage{physics}
\usepackage{comment}

\usepackage{soul}

\def\nb{N_\text{b}}

\def\Eh{\text{E}_\text{h}}
\def\eem{\text{e}} %natural number
\def\iim{\text{i}} %imaginary unit

\usepackage[dvipsnames]{xcolor}

\def\dd{\text{d}}
\def\Nbas{N_\text{b}}
\def\Nperm{N_\text{perm}}
\newcommand{\itfun}[1]{\mathcal{I}_{#1}}
\newcommand{\titfun}[1]{\tilde{\mathcal{I}}_{#1}} %asymptotic form

\makeatletter
\DeclareFontFamily{OMX}{MnSymbolE}{}
\DeclareSymbolFont{MnLargeSymbols}{OMX}{MnSymbolE}{m}{n}
\SetSymbolFont{MnLargeSymbols}{bold}{OMX}{MnSymbolE}{b}{n}
\DeclareFontShape{OMX}{MnSymbolE}{m}{n}{
    <-6>  MnSymbolE5
   <6-7>  MnSymbolE6
   <7-8>  MnSymbolE7
   <8-9>  MnSymbolE8
   <9-10> MnSymbolE9
  <10-12> MnSymbolE10
  <12->   MnSymbolE12
}{}
\DeclareFontShape{OMX}{MnSymbolE}{b}{n}{
    <-6>  MnSymbolE-Bold5
   <6-7>  MnSymbolE-Bold6
   <7-8>  MnSymbolE-Bold7
   <8-9>  MnSymbolE-Bold8
   <9-10> MnSymbolE-Bold9
  <10-12> MnSymbolE-Bold10
  <12->   MnSymbolE-Bold12
}{}

\let\llangle\@undefined
\let\rrangle\@undefined
\DeclareMathDelimiter{\llangle}{\mathopen}%
                     {MnLargeSymbols}{'164}{MnLargeSymbols}{'164}
\DeclareMathDelimiter{\rrangle}{\mathclose}%
                     {MnLargeSymbols}{'171}{MnLargeSymbols}{'171}
\makeatother
\usepackage{color}
\usepackage{multirow}

\usepackage[unicode]{hyperref}
\hypersetup{
   unicode=true,          % non-Latin characters in Acrobat??s bookmarks
   plainpages=false,
   colorlinks=true,       % false: boxed links; true: colored links
   citecolor=blue,        % color of links to bibliography
}

\usepackage{url}

\DeclareUnicodeCharacter{2212}{-}

\begin{document}

\title{
On the inclusion of cusp effects in expectation values with explicitly correlated Gaussians
}

\date{\today}% It is always \today, today,

\author{P\'eter Jeszenszki}
\affiliation{Institute of Chemistry, ELTE, Eötvös Loránd University, 
Pázmány Péter sétány 1/A, Budapest, H-1117, Hungary}
\author{Robbie T. Ireland}
\affiliation{Institute of Chemistry, ELTE, Eötvös Loránd University, 
Pázmány Péter sétány 1/A, Budapest, H-1117, Hungary}
\affiliation{School of Chemistry, University of Glasgow,
University Avenue, G12 8QQ, Glasgow, United Kingdom}
\author{D\'avid Ferenc} 
\author{Edit M\'atyus} 
\email{edit.matyus@ttk.elte.hu}
\affiliation{Institute of Chemistry, ELTE, Eötvös Loránd University, 
Pázmány Péter sétány 1/A, Budapest, H-1117, Hungary}

\begin{abstract}
\noindent %
This paper elaborates the integral transformation technique of [K. Pachucki, W. Cencek, and J.~Komasa, J.~Chem.~Phys. 122, 184101 (2005)] and uses it for the case of the non-relativistic kinetic and Coulomb potential energy operators, as well as for the relativistic mass-velocity and Darwin terms.
The techniques are tested for the ground electronic state of the helium atom and perturbative relativistic energies are reported for the ground electronic state of the H$_3^+$ molecular ion near its equilibrium structure. 
\end{abstract}

\maketitle

%\tableofcontents

\section{Introduction}
\noindent We wish to dedicate this paper to István Mayer's memory. Two of us attended his undergraduate special course (called `speci' among the students) at ELTE that he held until ca.~2010. During our everyday work, we still frequently point to simple calculations and theorems that we have learned from him and from his book \cite{mayer_simple_2003}. As students, and later, as young researchers, we got to know him as an infinitely patient and supportive person towards the youths and their small things in research.
His every reasoning and calculation was simple, because he made them simple and made every small step clear.
In this spirit, we work out in detail the theoretical background for a nice technique proposed by Pachucki, Cencek, and Komasa that makes it possible to correct for the effects of the missing cusp of Gaussian basis functions during the evaluation of the `singular' integrals in the Breit--Pauli Hamiltonian \cite{pachucki_acceleration_2005}. %
We imagine presenting this work on a research seminar:
we can almost see István Mayer sitting and smiling in the first row of the auditorium and he has several comments and questions. We wonder: what are they?

Pachucki, Cencek, and Komasa \cite{pachucki_acceleration_2005} proposed
the integral transformation technique to enhance the convergence of the expectation values of terms of the Breit--Pauli Hamiltonian that were known
to be difficult to evaluate precisely in the commonly used explicitly correlated Gaussian (ECG) basis sets \cite{jeziorski_high-accuracy_1979,cencek_manyelectron_1993,suzuki_stochastic_1998,mitroy_theory_2013}
\begin{align}
  \Theta_i(\br) 
  &= 
  \exp \left[%
    -\left( \br-\bs_i \right)^T \underline{\bA}_i \left( \br - \bs_i \right)
  \right] \ ,
  \label{eq:ECGansatz}
\end{align}
where $\br\in\mathbb{R}^{3n}$ is the position vector of the particles, while $\bs_i\in\mathbb{R}^{3n}$ and $\underline{\bA}_i=\bA_i\otimes 1^{[3]}$ with $\bA_i\in\mathbb{R}^{n\times n}$ are parameters of the basis function. The parametrization is selected by minimization of the non-relativistic energy.
The advantage of the ECG basis set is that it is an $n$-particle basis, for which 
analytic matrix elements can be derived for almost all physically relevant operators.
At the same time, it is also well-known that the Gaussian functions fail to reproduce the analytic properties of the exact non-relativistic wave function at the particle-particle coalescence points (cusps) and in the asymptotic range for large particle-particle separations. The integral transformation technique offers a possibility to correct for the missing cusp effects.

We start the present work by writing out the theoretical background of Ref.~\cite{pachucki_acceleration_2005} in detail. During this work, we have noticed that the ideas used for the `integral transform'  (IT) evaluation of the perturbative relativistic corrections may be more generally applicable. 
In a nutshell, instead of directly evaluating the expectation value
of some physical quantity with the approximate wave function 
\begin{align}
  \langle \hat{O} \rangle 
  =
  \int \dd r_1 \ldots \dd r_N\ \Psi(\br_1,\ldots,\br_N)^\ast \hat{O}\ \Psi(\br_1,\ldots,\br_N) \; ,
\end{align}
it becomes possible to incorporate the effects of the cusp of the exact wave function. 
An appropriate transformation is defined by introducing $\itfun{\hat{O}}$ with variable $\xi$, and the integral is calculated in two parts,
\begin{align}
  \llangle \hat{O}\rrangle
  =
  \int_{0}^{\xi_\Lambda}
    \dd \xi\ 
     \itfun{\hat{O}}(\xi)
  +
  \int_{\xi_\Lambda}^\infty 
    \dd \xi\ 
    \titfun{\hat{O}}(\xi) \; , \label{slseparation}
\end{align}
where we introduced the $\llangle \rrangle$ notation to emphasize the difference from the standard expectation value labelled with $\langle \rangle$.
In the short-range part, $0\leq\xi\leq \xi_\Lambda$, the cusp has a negligible effect and it can be accurately computed with an ECG basis. For the long-range part, $\xi_\Lambda<\xi<\infty$, the exact 
cusp condition can be incorporated in the asymptotic tail of the transformed function ($\titfun{\hat{O}}(\xi)$) by considering the analytic behaviour of the wave function near the coalescence points.

In Sections~\ref{sec:ITCD} and \ref{sec:FTkin}, we work out the theoretical background and the analytic form of the long-range integrand for two types of integral transforms. 
Section~\ref{sec:implem} is about the implementation, technical details and observations. 
Numerical results are presented for the relativistic calculations in Sec.~\ref{sec:rel}, for the non-relativistic calculations in Sec.~\ref{sec:nonrel}, and the paper ends with a summary and conclusions (Sec.~\ref{sec:summary}).

\section{Integral transform for the Coulomb interaction and the Dirac delta of the coordinate \label{sec:ITCD}}
\noindent In this section, we will consider the inclusion of the cusp effect for 
spatial integrals of operators that can be related to the inverse of the particle-particle distance, $1/r$. So, let's first consider the interaction between an electron and a nucleus, which is fixed at the origin. In the matrix-element calculations, the relationship below is commonly used during the evaluation of the Coulomb integrals with Gaussian orbitals \cite{boys_electronic_1950,helgaker_molecular_2008}
\begin{align}
  \frac{1}{r_i} = \frac{2}{\sqrt{\pi}} \int_0^\infty \dd t\ \eem^{-r_i^2 t^2} 
  \label{eq:coulomb} \ ,
\end{align}
where the index $i$ indicates the index of the electron.
This relation can be understood as an integral transform (we call it $t$-transform) generation of $1/r_i$.
Furthermore, by using 
\begin{align}
  -4\pi \delta(\br_i)  = \Delta_{\br_i} \frac{1}{r_i} \; ,
\end{align}
we can write, following Ref.~\cite{pachucki_acceleration_2005},
\begin{align}
  \delta(\br_i)  
  =
  -\frac{1}{2\pi^{3/2}} 
  \int_{0}^\infty \dd t\ 2t^2(3-2t^2r_i^2)\ \eem^{-r^2_i t^2} \; .
  \label{eq:ddelta}
\end{align}
So, both operators can be generated by a $t$-integral
\begin{align}
  F(\bos{r}_i)
  =
  \int_0^\infty \dd t\ f(\bos{r}_i,t)\ \eem^{-r_i^2 t^2} \;,
\end{align}
where
\begin{align}
\text{for}\quad F(\bos{r}_i)=1/r_i:\quad
  f(\bos{r}_i,t) = 2/\sqrt{\pi}\; ,
\end{align}
and 
\begin{align}
\text{for}\quad F(\bos{r}_i)=\delta(\br_i):\quad
  f(\bos{r}_i,t) 
  = 
  -\pi^{-\frac{3}{2}}
  t^2 (3-2t^2r_i^2)\; .
\end{align}
Then, by generalizing Pachucki, Cencek, and Komasa's work for $\delta(\br)$ \cite{pachucki_acceleration_2005},
we re-write the expectation value for $F(\br_i)$ as
\begin{align}
  \langle%
    \Psi| F(\bos{r}_i)| \Psi
  \rangle
  &=
  \int \dd\br_1 \ldots \dd\br_N\ 
    \psi(\br_1,\ldots,\br_N)^\ast\ 
    F(\bos{r}_i)\ \psi(\br_1,\ldots,\br_N) \nonumber \\
  &=
  \int \dd\br_1 \ldots \dd\br_N\ 
    \left[\int_0^\infty \dd t\ 
    f(\bos{r}_i,t)\ \eem^{-r_i^2t^2}\right] |\psi(\br_1,\ldots,\br_N)|^2 \nonumber \\
  &=
  \int_0^\infty \dd t\ \int \dd\br_1 \ldots \dd\br_N\ 
    f(\bos{r}_i,t)\ \eem^{-r_i^2t^2} |\psi(\br_1,\ldots,\br_N)|^2 \nonumber \\
  &=
  \frac{1}{N} \int_0^\infty \dd t\ \int \dd\br_i\  
    f(\bos{r}_i,t)\ \eem^{-r_i^2t^2} \rho(\br_i) \label{Fexpwithdens}%\\
\end{align}
where $\rho(\br_i)$ is the one-electron density function,
\begin{align}
    \rho(\br_i) &=  N \int \left( \prod_{\stackrel{j =1}{ j \ne i} }^N \dd\br_j \right)
     |\psi(\br_1,\ldots,\br_N)|^2 \ .
     \label{eq:density}
\end{align}
Next, we define the integral transform function for $F(\br_i)$
as 
\begin{align}
  \label{eq:itf} \itfun{F(\br_i)}(t)
  &= \frac{1}{N}
  \int \dd\br_i\  
    f(\bos{r}_i,t)\ \eem^{-r_i^2t^2} \rho(\br_i) \ ,
\end{align}
which can be substituted back into \rref{Fexpwithdens},
\begin{align}
     \langle%
    \Psi| F(\br_i)| \Psi
  \rangle
  &= \int_0^\infty \dd t\ 
    \itfun{F(\br_i)}(t)  \; . \label{eq:fullintF}  
\end{align}
 The integral $\itfun{F(\br_i)}(t)$ can be written in an analytic form for `any' polynomial $f(\br_i,t)$ of $r_i$ and $t$. 
In particular, 
\begin{align}
  \text{for}\ F(\br_i)=1/r_i: \quad 
  \itfun{1/r_i}(t)
  =
  \frac{2}{\sqrt{\pi} N}
  \int \dd\br_i\  
    \eem^{-r_i^2t^2} \rho(\br_i)
    \label{eq:intfcoulomb}    
\end{align}
and
\begin{align}
  \text{for}\ F(\br_i)=\delta(\br_i): \quad
  \itfun{\delta(\br_i)}(t)
  =
  -\frac{1}{\pi^{\frac{3}{2}} N}
  \int \dd\br_i\  
    t^2 (3-2t^2r_i^2)\ 
    \eem^{-r_i^2t^2} \rho(\br_i) \; .
    \label{eq:intfdelta}
\end{align}
At first sight, it may seem strange that we introduce these complicated integral expressions, Eqs.~(\ref{eq:itf})--(\ref{eq:intfdelta}). This is especially true for the integral of Dirac delta that could be immediately obtained from the density at the origin. But, it is  difficult to calculate the density at this point, due to the cusp of the wave function. In numerical computations, $\rho(\br_i)$ is expanded in terms of a finite number of basis functions. The commonly used Gaussian functions are smooth everywhere and they miss the correct description of the cusp \cite{helgaker_molecular_2008,suzuki_stochastic_1998,hattig_explicitly_2012}. 

The integral transformation in \rref{eq:itf} widens out the effect of the density to a finite interval due to the term $\eem^{-r_i^2t^2}$ (for finite $t$ values), and over this finite interval, the density can be represented accurately with smooth functions. 
The original integral value is obtained by integration for $t\in[0,+\infty)$. 
For larger $t$ values, the Gaussian in \rref{eq:itf} becomes narrower and makes the short-range contribution (cusp) more important to $\itfun{F(\br_i)}(t)$. 

In the following paragraphs, it will be shown that for large $t$ values,
the analytic form of the integrand can be deduced from analytic properties of the density near the cusp.
To be able to incorporate these analytic results, the full integral is evaluated as the sum of a short-range, $t\in[0,t_\Lambda]$, and a long-range, $t\in[t_\Lambda,\infty)$, part {(\rref{slseparation})}:
\begin{align}
  \left\llangle \frac{1}{r_i} \right\rrangle 
  =
  \int_0^{t_\Lambda} \itfun{1/r_i}(t) \dd t 
  +
  \int_{t_\Lambda}^\infty \titfun{1/r_i}(t) \dd t \; . \label{eq:1orslsep}
\end{align}
The short-range part is evaluated by direct integration over the finite interval $t\in[0,t_\Lambda]$ (Appendix~\ref{sec:Gint}).
To calculate the long-range part including the cusp effects, the following considerations are necessary.

\subsection{Derivation of the long-range part from the cusp condition\label{sec:longrange}}
\noindent According to Kato's cusp condition \cite{katoEigenfunctionsManyparticleSystems1957a,mayer_simple_2003,helgaker_molecular_2008}, the following relations hold for the exact non-relativistic wave function (in Hartree atomic units) for the electron-nucleus and for the electron-electron
coalescence points, respectively,
\begin{align}
  \lim_{r_{iA}\rightarrow 0}
  \left \langle \frac{%
    \partial  \Psi 
  }{%
    \partial {r_{iA}} }\right \rangle_{\vartheta,\varphi}
  =
  -Z_A\psi(\bos{r}_{iA}=\bos{0})
  \quad\text{and}\quad
  \lim_{r_{ij}\rightarrow 0}
  \left \langle \frac{%
    \partial \Psi
  }{%
    \partial {r_{ij}}
  } \right \rangle_{\vartheta,\varphi}
  =\frac{1}{2}\psi(\bos{r}_{ij}=\bos{0}) \ , \label{Katocusp}
\end{align}
where $\langle \rangle_{\vartheta,\varphi}$ indicates averaging for the spherical angles, $Z_A$ is the nuclear charge number, $r_{iA}$ is the distance between electron $i$ and nucleus $A$, and $r_{ij}$ is the distance between electrons $i$ and $j$. These conditions are valid only if the wave function does not have a node at the coalescence point, otherwise, higher derivatives must be considered for a good description of the wave function in this regime \cite{packCuspConditionsMolecular1966,kutzelnigg_theory_1994}. The coalescence condition can be further elaborated by considering the effect of higher derivatives of the wave function \cite{rassolovBehaviorElectronicWave1996,tewSecondOrderCoalescence2008}, which can be also affected by three-particle coalescence conditions \cite{fournaisSharpRegularityResults2005,myersFockExpansionKato1991}. In this paper, we use the simplest, original conditions of \rref{Katocusp} that give the following relations 
\cite{steinerChargeDensitiesAtoms1963}:
\begin{align}
  \lim_{r_{iA}\rightarrow 0}
  \frac{%
    \partial {\left \langle \rho \right \rangle_{\vartheta,\varphi}}
  }{%
    \partial {r_{iA}}
  }
  =
  -2 Z_A\rho(\bos{0})
  \quad\text{and}\quad
  \lim_{r_{ij}\rightarrow 0}
  \frac{%
    \partial {\left \langle \eta \right \rangle_{\vartheta,\varphi}}
  }{%
    \partial {r_{ij}}
  }=\eta(\bos{0}) \; ,
  \label{CuspDens}
\end{align}
where $\rho$ is the one-electron density, Eq.~(\ref{eq:density}), and $\eta$ labels
the pair correlation function \cite{kimballShortrangeCorrelationsStructure1975},
\begin{align}
  \eta(\br)
  =
  N(N-1) \int \left(\prod_{k=2}^N \dd\br_k \right)
     |\psi(\br_2 + \br, \br_2, \br_3, \br_4, \ldots,\br_N)|^2 \ .
\end{align}
that can also be understood also as a  quantity proportional to the pseudo-particle density corresponding to the relative motion for a pair of electrons \cite{suzuki_stochastic_1998}.

Then, we may consider the expansion of the spherically averaged  density and pair correlation function by the coalescence point taken as the origin ($\bos{0}$):
\begin{align}
    \langle \rho\rangle_{\vartheta,\phi} (r) 
    &= 
      \rho(\bos{0}) 
      - 2 Z_A\rho(\bos{0}) r 
      + \sum_{j=2}^m B_j r^{j} + \mathcal{O}(r^{m+1}) 
      \ , \label{eq:Taylorrho} \\
    \langle \eta\rangle_{\vartheta,\phi}(r)  
    &= 
    \eta(\bos{0}) 
      + \eta(\bos{0}) r 
      + \sum_{j=2}^m B_j r^{j} + \mathcal{O}(r^{m+1}) 
    \ .
    \label{eq:Tayloreta} 
\end{align}
To obtain the asymptotic form of  $\itfun{F(\br_i)}$, labelled with
$\titfun{F(\br_i)}$ (where tilde refers to the fact that it is valid for the asymptotic range),
we insert the density expansion, \rref{eq:Taylorrho}, in the definition of the 
integral transform function, \rref{eq:itf} and integrate out the angular coordinates. We explicitly show  the calculation for $\titfun{F(\br_i)}$ (and it can be carried out analogously for $\titfun{F(\br_{ij})}$ using \rref{eq:Tayloreta})  
\begin{align}
\text{for }t>t_\Lambda:\quad  \nonumber  \\
  \titfun{F(\br_i)}(t)
  &=
  \frac{1}{N} \int_0^{2\pi}\dd\phi\int_{-1}^1\dd(\cos\vartheta) \int_0^\infty \dd r_i\ r_i^2\ f(\br_i,t)\ \eem^{-r_i^2t^2} 
    \rho(\bos{r}_i) 
  \nonumber \\
  &=  
  \frac{1}{N} \int_0^\infty \dd r_i\ r_i^2\ f(\br_i,t)\ \eem^{-r_i^2t^2} 
    4\pi\langle\rho \rangle_{\vartheta,\phi}(r_i)
  \nonumber \\
  &=  
  \frac{4\pi}{N} \int_0^\infty \dd r_i\ r_i^2\ f(\br_i,t)\ \eem^{-r_i^2t^2} 
  \left[%
    \rho(\bos{0}) 
    - 2 Z_A\rho(\bos{0}) r_i 
    + \sum_{j=2}^m B_j r_i^{j} + \mathcal{O}(r_i^{m+1})    
  \right] \; ,
\end{align}
where for practical reasons, we truncate the expansion after some (`appropriate') $m$ value.
The one-dimensional integral for $r_i$ can be evaluated by analytic or numerical integration. 
For $F(\br_i)=1/r_i$ with $f(\br_i,t)=2 \pi^{-1/2}$ in Eq.~(\ref{eq:coulomb}), we obtain the asymptotic form as
\begin{align}
\text{for }t>t_\Lambda:\quad
  \titfun{1/r_i}(t)
  =
  \frac{1}{t^3 N}\left(%%
  2\pi\rho(\bos{0})
  -8\sqrt{\pi} Z_A \rho(\bos{0}) \frac{1}{t}
  +\sum_{j=2}^m B^{[1/r_i]}_j \frac{1}{t^{j}}
  \right) \; ,
  \label{eq:asymcoulomb}  
\end{align}
while for $F(\br_i)=\delta(\br_i)$, $f(\br_i,t)=-t^2(3-2t^2r_i^2) \pi^{-3/2}$ in Eq.~(\ref{eq:ddelta}), we have
\begin{align}
\text{for }t>t_\Lambda:\quad
  \titfun{\delta(\br_i)}(t)
  =
  \frac{1}{t^2 {N}} 
  \left( \frac{4Z_A \rho(\bos{0})}{\sqrt{\pi}}  - 
  \frac{2}{\sqrt{\pi}}
    \sum_{j=2}^m
      B^{[\delta(\br_i)]}_j
      \frac{%
        1
        }{ 
        t^{j-1}
        } 
        \right)  \, .
        \label{eq:asymdelta}
\end{align}
It is interesting to note that the asymptotic tail of the Coulomb interaction, Eq.~(\ref{eq:asymcoulomb}), decays faster than that of the Dirac delta, Eq.~(\ref{eq:asymdelta}), leading to a faster convergence in a finite basis representation.
Using \rrefsa{eq:asymcoulomb} and \rrefsb{eq:asymdelta}, the integral from $t_\Lambda$ to $\infty$ is obtained in an analytic form as
\begin{align}
     \int_{t_\Lambda}^\infty \titfun{1/r_i}(t)\ \dd t  \ &= 
     \frac{1}{t_\Lambda^2 {N}}\left(%%
  \pi\rho(\bos{0})
  -\frac{8}{3}\sqrt{\pi} Z_A \rho(\bos{0}) \frac{1}{t_\Lambda}
  +\sum_{j=2}^m \frac{B^{[1/r_i]}_j}{j+2} \frac{1}{t_\Lambda^{j}}
  \right) \ , \label{intIri} \\
  \int_{t_\Lambda}^\infty \titfun{\delta(\br_i)}(t)\ \dd t  \ &= 
    \frac{1}{t_\Lambda {N}} 
  \left( \frac{4Z_A \rho(\bos{0})}{\sqrt{\pi} }  - 
  \frac{2}{\sqrt{\pi}}
    \sum_{j=2}^m
     \frac{ B^{[\delta(\br_i)]}_j}{j}
      \frac{%
        1
        }{ 
        t_\Lambda^{j-1}
        } 
        \right) \; . \label{intIdeltari}
\end{align}
 
Although both expressions contain the particle density at the coalescence point, $\rho(\bos{0})/N=\langle \Psi|\delta(\br_i)|\Psi\rangle$, that is inaccurately represented in a(n explicitly correlated) Gaussian basis, we can obtain its precise value by using:
\begin{align}
\frac{1}{N}\rho(\bos{0})
  =
  \llangle \delta(\br_i) \rrangle
  =
  \int_{0}^{t_\Lambda} \itfun{\delta(\br_i)}(t)\ \dd t
  +
  \int_{t_\Lambda}^\infty \titfun{\delta(\br_i)}(t)\ \dd t
  \label{eq:ITdelta}
\end{align}
in an iterative procedure. 
First, the short-range integral (first term in the right-hand side of Eq.~(\ref{eq:ITdelta})) is calculated by a one-dimensional quadrature (since this integrand is too complicated for an analytic evaluation),
while the $B_j$ parameters in the long-range part are obtained by fitting the asymptotic part, Eq.~(\ref{eq:asymdelta}), to data points. The data set for the fit corresponds to `intermediate'-range $t$ values (for practical details, see Secs.~\ref{sec:rel}, \ref{sec:nonrel}, and Appendix \ref{sec:rhoeta}).
Then, using $\rho(\bos{0})$, obtained directly from numerical integration, the $\titfun{\delta(\br_i)}(t)$ asymptotic function can be evaluated. In the last step, we calculate the integrals in Eq.~(\ref{eq:ITdelta}) that results in an improved value for
$\rho(\bos{0})$ and an improved $\titfun{\delta(\br_i)}(t)$ asymptotic form. The iteration converges in a few cycles as it was noted already in Ref.~\cite{pachucki_acceleration_2005}.
Once we have the precise value for $\rho(\bos{0})$, we can have a good representation for the asymptotic tail of the Coulomb interaction, 
$\titfun{1/r_i}(t)$ in Eq.~(\ref{eq:asymcoulomb}). Then, the integral value for the Coulomb interaction including also the cusp effect can be obtained as:
\begin{align}
  \left\llangle \frac{1}{r_i} \right\rrangle 
  =
  \int_0^{t_\Lambda}
  \itfun{1/r_i}(t)\ \dd t 
  +
  \int_{t_\Lambda}^\infty \titfun{1/r_i}(t)\ \dd t \; .
\end{align}

For computing $\llangle \delta(\br_{ij}) \rrangle$ and $\left\llangle 1/r_{ij}  \right\rrangle$ a similar approach is used, but it is necessary to substitute $\rho(\bos{0})$, $Z_A$, and $N$ with $\eta(\bos{0})$,  $-1/2$, and $N(N-1)$, respectively in %\rrefsa{intIri} 
Eqs. (\ref{eq:asymcoulomb})--(\ref{intIdeltari}). The final working equations are 
\begin{align}
\hspace{-3cm} \text{for }t>t_\Lambda:\ \hspace{2cm}& \nonumber \\
  \titfun{1/r_{ij}}(t)
  &=
  \frac{1}{t^3 N(N-1)}\left(%%
  2\pi\eta(\bos{0})
  +4\sqrt{\pi} \eta(\bos{0}) \frac{1}{t}
  +\sum_{k=2}^m B^{[1/r_{ij}]}_k \frac{1}{t^{k}}
  \right) \; ,
  \label{eq:asymcoulombrij} \\
  \titfun{\delta(\br_{ij})}(t)
  &=
  -\frac{1}{t^2 {N(N-1)}} 
  \left( \frac{2\eta(\bos{0})}{\sqrt{\pi}}  + 
  \frac{2}{\sqrt{\pi}}
    \sum_{k=2}^m
      B^{[\delta(\br_{ij})]}_k
      \frac{%
        1
        }{ 
        t^{k-1}
        } 
        \right) \ , 
        \label{eq:asymdeltarij}
\end{align}
and
\begin{align}
   \int_{t_\Lambda}^\infty 
    \titfun{1/r_{ij}}(t)\ \dd t  \ 
   &= 
  \frac{1}{t_\Lambda^2 N(N-1)}
  \left(%%
    \pi\eta(\bos{0})
    +\frac{4}{3}\sqrt{\pi} \eta(\bos{0}) \frac{1}{t_\Lambda}
    +\sum_{k=2}^m \frac{B^{[1/r_{ij}]}_k}{k+2} \frac{1}{t_\Lambda^{k}}
   \right) \ , \label{intIrij} \\
   \int_{t_\Lambda}^\infty \titfun{\delta(\br_{ij})}(t)\ \dd t  \ &= 
  \frac{1}{t_\Lambda N(N-1)} 
  \left( 
  -\frac{2 \eta(\bos{0})}{\sqrt{\pi} }  
  -\frac{2}{\sqrt{\pi}}
    \sum_{k=2}^m
     \frac{ B^{[\delta(\br_{ij})]}_k}{k}
      \frac{%
        1
        }{ 
        t_\Lambda^{k-1}
        } 
        \right) \; , \label{intIdeltarij}
\end{align}
with 
\begin{align}
  \frac{1}{N(N-1)} \eta(\bos{0}) 
  = 
  \llangle
    \delta(\bos{r}_{ij}) \label{eq:ITdeltarij}
  \rrangle \; ,
\end{align}
where the precise value of $\llangle\delta(\bos{r}_{ij}) \rrangle$ is obtained in an iterative procedure,  similarly to $\llangle\delta(\bos{r}_{iA}) \rrangle$.

\section{Fourier transform for the kinetic energy and the mass-velocity terms \label{sec:FTkin}} 
\noindent To calculate integrals of momentum operators, it is convenient to switch to momentum space. 
The Fourier transform of an ECG preserves the mathematical form of the function, and we need to consider only the changes in the parameterization.
So, the Fourier transform of the basis function in \rref{eq:ECGansatz}
is \cite{pachucki_acceleration_2005} 
\begin{align}
  \bar{\Theta}_i(\bp)
  =
  |\bA|^{-\frac{3}{2}} 
  \exp\left[%
    -(\bp-\bar{\bs}_i){^\text{T}} \bar{\ubA}_i (\bp-\bar{\bs}_i)+\bar{C}_i
  \right] \; , 
  \label{eq:FTecg}
\end{align}
with
$\bar{\bs}_i=-2\iim\bs_i \ubA_i$,
$\bar{\bA}_i=\frac{1}{4}\bA_i^{-1}$,
and $\bar{C}_i=-\bs_i^\text{T}\ubA_i\bs_i$.
For the expectation value of the non-relativistic kinetic ($k=2$)
and of the mass-velocity ($k=4$) operators, we have to evaluate
\begin{align}
  \langle \Psi | p_1^k |\Psi\rangle
  &=
  \int \dd \bp_1\ldots \dd \bp_n\ p_1^k\  |\bar{\Psi}(\bp_1,\ldots,\bp_n)|^2 \nonumber \\
  &=
 \frac{1}{N}\int \dd \bp_1\ p_1^k\ \bar{\rho}(\bp_1) { \ ,}
\end{align}
where $\bar{\rho}(\bp_1)$ is the momentum density function. The angular part of the integral can be evaluated according to  Eq.~(\ref{eq:pangint}),
\begin{align}
  \langle \Psi | p_1^k |\Psi\rangle &=
  \frac{4\pi}{N} \int_0^\infty \dd p\ p^{k+2}\ 
 \left \langle \bar{\rho} \right \rangle_{\vartheta,\varphi}(p) =
  \int_0^\infty \dd p\ \itfun{p^k}(p) \; ,\label{pkexpval} 
\end{align}
where $\left \langle \bar{\rho} \right \rangle_{\vartheta,\varphi}(p)$ labels 
the spherically averaged momentum density.
The explicit integrals for $k=2$ ($p^2$) are evaluated in  Eqs.~(\ref{eq:ptwo})--(\ref{eq:ptwos}), and the calculation can be, in principle, carried out similarly for $k=4$, but we used quadrature integration, because it was fast and sufficiently accurate (Sec.~\ref{sec:rel}).
It is interesting to note that the momentum density is spherically symmetric (second step in \rref{pkexpval}), even if $\bos{s}_i\neq0$. This observation is connected with the properties of the Fourier transform of the ECG, \rref{eq:FTecg}, in which any coordinate-space shift vector appears as purely imaginary vector. 

Similarly to the $t$-transform (Sec.~\ref{sec:ITCD}), the cusp dominating the small-scale behavior in coordinate space is important for the long-range part in the inverse (now momentum) space. 
To be able to exploit the different characteristics for the two ranges (short and long), the integral is evaluated in two parts,
\begin{align}
  \llangle 
    \Psi | p_1^k |\Psi
  \rrangle
  =
  %\sum_{i=0}^{p_\Lambda} w_i \itfun{p^k}(p_i)
  \int_{0}^{p_\Lambda} \dd p\ \itfun{p^k}(p)
  +
  \int_{p_\Lambda}^\infty \dd p\ \titfun{p^k}(p)\; ,
  \label{eq:pkshl}
\end{align}
where the short-range part corresponds to the first term and is calculated from the ECG representation by direct integration up to some appropriate $p_\Lambda$ threshold. 
The long-range part (second term) is determined by the cusp effects, and its analytic properties can be derived for the asymptotic tail. We will label this analytic asymptotic expression by $\titfun{p^k}(p)$ that is derived in the forthcoming subsection.

\subsection{The asymptotic tail of the momentum density \label{sec:asympdens}}
To show the connection of the short-range behaviour in coordinate space dominated by the particle-particle coalescence point(s) and the long-range behaviour in momentum space, we need to consider a common theorem from numerical analysis \cite{mercier_introduction_2014} which connects the smoothness of a function, $f(x)$, with the asymptotic behavior after Fourier transformation, $\tilde{f}(k) = \int f(x) \eem^{ikx} \dd x$. The smoothness of $f(x)$ is defined by the number of continuous derivatives. If $f(x)$ is infinitely differentiable, $f(x) \in C^\infty$, or in other words $f(x)$ is  smooth, then $\tilde{f}(k)$ decays exponentially fast at large $k$ values. If the $n$th derivative corresponds to a Dirac delta function,  then the first $n-2$ derivatives are continuous, $f(x) \in C^{n-2}$, and $\tilde{f}(k)$ decays polynomially with $1/k^{n}$ 
(Appendix~\ref{sec:smoothness}). 

We use this theorem, following Ref.~\cite{kimballShortrangeCorrelationsStructure1975}, to determine the analytic form for the asymptotic tail of the momentum density function, $\bar{\rho}(\bp)$ 
\begin{align}
    \bar{\rho}( {\bp} ) &= \frac{1}{8 \pi^3} \int  \, \eem^{\iim \bp (\br -\br')} \Gamma(\br', \br)\ \dd \br\ \dd \br' \  \label{momdens} \\
\end{align}
with the one-particle density matrix, 
\begin{align}
    \Gamma(\br', \br) &= {N}\int \Psi^*(\br', \br_2, \dots, \br_{N})  \Psi(\br, \br_2, \dots, \br_{N}) \prod_{i=2}^{N} \dd \br_i  \ . \label{densitymat}
\end{align}
By substituting \rref{densitymat} into \rref{momdens} and by exchanging the order of integration, we arrive at an alternative expression for the momentum density,
\begin{align}
     \bar{\rho}( {\bp} ) &=  {N}\int \widetilde{\Psi}^*(\bp, \br_2, \dots, \br_{N})  \widetilde{\Psi}(\bp, \br_2, \dots, \br_{N}) \prod_{i=2}^{N} \dd \br_i \,  \ , \label{momdens2} \\
    \widetilde{\Psi}(\bp,\br_2,\dots, \br_{N}) & = \frac{1}{\sqrt{8 \pi^3}}  \int  \, \eem^{-\iim \bp \br }  \Psi(\br,\br_2,\dots, \br_{N})\ \dd \br \ . \label{momwf0}
\end{align}
To describe the asymptotic tail in momentum space, it is sufficient to consider those regions of the wave function for which the singularity occurs for higher-order derivatives (Appendix~\ref{sec:smoothness}). 
These regions are the points at the position of the nuclei and at the electron-electron coalescence points, where the exact wave function cusps. 

Let us focus on a cusp at nucleus $A$ located at $\bR_A$. Then, 
we consider the integral form of the cusp condition \cite{bingelBehaviourFirstorderDensity1963,packCuspConditionsMolecular1966}, 
\begin{align}
 \text{for }\bos{r}_i \approx \bos{R}_A: 
\quad \Psi(\br_1, \br_2, \dots, \br_i, \dots, \br_{N} )  
  \approx
 \left\{ 1-\left[Z_A+f_A(\vartheta_i,\varphi_i) \right]\sqrt{(\bos{r}_i-\bos{R}_A)^2}\right\} \Phi_i(\br_1, \br_2, \dots, \br_i, \dots, \br_{N} ) \ ,   \label{cuspforwf}
\end{align}
where $Z_A$ is the nuclear charge number, $f_A(\vartheta_i,\varphi_i)$ is an angular term, and $\Phi(\br_1 ,\dots \br_{N})$ is a continuous function at least up to its second derivative according to $\br_i$ at $\bos{R}_A$ for every particle $i$. The $f_A(\vartheta_i,\varphi_i)$ term accounts for the angular dependence (that is not generally spherically symmetric) of the wave function around the cusp. A more specific form for $f_A(\vartheta_i,\varphi_i)$ can be obtained, if we consider the expansion of the one-electron contribution of the wave function around the cusp using the eigenfunctions of the hydrogen atom \cite{bingelBehaviourFirstorderDensity1963,packCuspConditionsMolecular1966,mayer_simple_2003}. As the leading-order contribution of the radial part is related to $r^\ell$, where $\ell$ is the angular momentum quantum number, we can neglect all $\ell \ge 2$ angular terms  for the small $r$. 
So, to describe the non-spherical angular dependence, it is sufficient to consider the linear combinations of the first-order spherical harmonics ($Y_{1m}(\vartheta,\varphi)$, $m=-1,0,1$).
It is shown in Appendix~\ref{sec:FTrf} that the angular dependence does not have an effect on  the large-momentum tail (that corresponds to the short $r$ range) \cite{bingelBehaviourFirstorderDensity1963,packCuspConditionsMolecular1966,mayer_simple_2003}. 

In order to examine the non-smoothness of the cusp, let us consider $\nabla_i^4 \Psi$ (that is commonly understood as $\nabla_i^4 \Psi=(\bos{\nabla}_i \cdot \bos{\nabla}_i)^2 \Psi$):
\begin{align}
  \nabla_i^4  \Psi  
  =&  
  -\left[\nabla_i^4  Z_A  \sqrt{(\bos{r}_i-\bos{R}_A)^2} \right] \Phi_i + \phi_i  \; ,
\label{nabla4onwf}
\end{align}
where $\phi_i$ collects remainder terms that are smooth with respect to $\br_i$ near $\bR_A$. (The effect from cusps due to the other particles  can be accounted for by summing up the contributions.)
If ${\nabla_i^4}$ acts on the cusp, a Dirac delta singularity appears,
\begin{align}
    -Z_A{\nabla_i^4}  \sqrt{(\bos{r}_i-\bos{R}_A)^2}
    =
    -Z_A\nabla_i^2 \frac{2}{\sqrt{(\bos{r}_i-\bos{R}_A)^2}} 
    &= 
    8 \pi Z_A \delta\left(\bos{r}_i-\bos{R}_A \right) \ . 
    \label{diracdelta}
\end{align}
For the fourth derivative, the integral in \rref{momwf0} can be evaluated using the properties of Dirac delta in \rref{diracdelta}, and thus, we obtain the leading-order contribution for large momentum,
\begin{align}
\text{for }|\bos{p}_i|>p_\Lambda: 
\quad \widetilde{\Psi}(\bos{r}_1, \dots, \bos{p}_i, \dots, \bos{r}_{N})  = 
    \frac{ 2\sqrt{2}  Z_A }{ \sqrt{\pi} p_i^4} 
    \Psi(\bos{r}_1, \bos{r}_2, \dots , \bos{R}_A, \dots, \bos{r}_{N})
    \eem^{-\iim{\bf p}_i \bos{R}_A} + \mathcal{O}\left( p_i^{-6} \right)
    \; .
    \label{eq:longpnuc}
\end{align}
This short calculation demonstrates that it is indeed the cusp that determines the large-momentum behaviour. In \rref{eq:longpnuc} (valid for large $p$), the next leading order comes with $p_i^{-6}$. Although the $p_i^{-5}$ term can be neglected in the asymptotic tail, further odd powers of $1/p_i$ are retained to account for (possible) higher-order singularities in the wave function \cite{rassolovBehaviorElectronicWave1996,tewSecondOrderCoalescence2008,fournaisSharpRegularityResults2005}.

To generalize the calculation to several nuclei and electrons, we consider the following Ansatz which includes the effect of all the cusps of the exact wave function of the many-particle system,
\begin{align}
  \Psi(\br_1, \br_2, \dots, \br_i, \dots, \br_{N} )  
  = 
 \sum_{i=1}^{N}
 %{\Bigg \{}
   {\bigg \{} 1-\sum_{A=1}^{N_\mathrm{nucl}}&
   \left[Z_A +f_A(\vartheta_i,\varphi_i) \right]
   \sqrt{(\bos{r}_i-\bos{R}_A)^2} \nonumber \\
 &+ 
 \sum_{j \ne i}^{N} 
   \left[\frac{1}{2} +g(\vartheta_{ij},\varphi_{ij}) \right]
   \sqrt{(\bos{r}_i-\bos{r}_j)^2} {\bigg \}} 
   \Phi_i(\br_1, \br_2, \dots, \br_i, \dots, \br_{N} ) 
 \ ,
\end{align}
where $g(\vartheta_{ij},\varphi_{ij})$ takes into account the angular dependence of the short-range electron-electron correlation,
similarly to the $f_A(\vartheta_i,\varphi_i)$ term for the electron-nucleus
cusp \cite{bingelBehaviourFirstorderDensity1963,packCuspConditionsMolecular1966}. 
The calculation of the large-momentum effect of the electron-electron coalescence can be carried out in a similar manner to the electron-nucleus case, Eqs.~(\ref{cuspforwf})--(\ref{eq:longpnuc}),
after a coordinate transformation to the center-of-mass and relative motion
coordinates including the $\br_{ij}$ displacement vector.
The same arguments apply for the electron-electron cusp as for the 
electron-nucleus case, with the only difference that the $Z_A$ nuclear charge number is replaced with $Z_\text{ee}=-1/2$.
Then, the asymptotic tail in momentum space for a many-electron-many-nucleus system is obtained as
\begin{align}
 \text{for }|\bos{p}_i|>p_\Lambda:  
\quad \bar{\Psi}(\bos{r}_1, \dots, \bos{p}_i, \dots, \bos{r}_{N})  =
    \frac{ \sqrt{2}  }{ \sqrt{\pi}p_i^4} \Bigg[ 2\sum_{A=1}^{ N_\mathrm{nucl}} Z_A &  
    \Psi(\bos{r}_1, \bos{r}_2, \dots , \bos{R}_A, \dots, \bos{r}_{N})
    \eem^{-\iim\bos{p}_i \bos{R}_A} \nonumber \\
    & - \sum_{j \ne i}^{{N}}   \Psi(\bos{r}_1, \bos{r}_2, \dots , \bos{r}_j, \dots, \bos{r}_{N})
    \eem^{-\iim\bos{p}_i \bos{r}_j}  \Bigg] +\mathcal{O}\left( p_i^{-6} \right)
    \; . \label{momwf}
\end{align}
To obtain the asymptotic tail for the momentum density, we substitute \rref{momwf} into \rref{momdens2},
\begin{align}
  &\text{for }|\bos{p}|>p_\Lambda:  \nonumber \\
  &\quad\bar{\rho}(\bos{p}) 
  = 
  \frac{2}{ \pi p^8} 
  \Bigg[ %
   4 \sum_{A=1}^{N_\mathrm{nucl}} Z_A^2 \, \rho(\bos{R}_A) 
   + 
   4 \sum_{A=1}^{N_\mathrm{nucl}} \sum_{B \ne A}^{N_\mathrm{nucl}}  
     Z_A Z_B \cos\left[ \bos{p} \left( \bos{R}_A-\bos{R}_B\right)\right] \Gamma(\bos{R}_A,\bos{R}_B) \nonumber \\
   &\hspace{2.2cm} 
   -2(N-1) \sum_{A=1}^{N_\mathrm{nucl}}  
   Z_A \left(%
     \int \eem^{\iim \bos{p} \left( \bos{r}_2-\bos{R}_A\right)} \Psi^*(\bos{R}_A, \bos{r}_2,  \dots, \bos{r}_{N})  \Psi(\bos{r}_2, \bos{r}_2,  \dots, \bos{r}_{N})  \prod_{i=2}^{N} \dd \br_i + \text{cc.} 
     \right)  \nonumber \\
   & \hspace{12cm} + 
     \eta(\bos{0}) 
   \Bigg]
    + \mathcal{O}\left( p^{-10}\right) \ ,\label{eq:rhopvec}
\end{align}
where `+cc.' means complex conjugation of the first term in the parenthesis.
The interesting result that the pair correlation function appears in the momentum distribution was first noticed in Refs.~\cite{kimballShortrangeCorrelationsStructure1975,yasuharaNoteMomentumDistribution1976}. Moreover, it was also found that it leads to a fifth-order cusp in the off-diagonal density matrix in the jellium model \cite{marchFormFirstorderDensity1975}. This fifth-order cusp has been derived recently for general atoms and molecules without using the known results from the momentum distribution \cite{cioslowskiOffdiagonalDerivativeDiscontinuities2020}, hence, the asymptotic tail in \rref{eq:rhopvec} can be obtained (as an alternative route to the present one) by Fourier-transforming the cusp condition of the off-diagonal density matrix (Eq.~(20) in Ref.~\cite{cioslowskiOffdiagonalDerivativeDiscontinuities2020}).

Furthermore, it can be shown by partial integration that 
that the integral term in \rref{eq:rhopvec} is proportional to $1/p^4$ for high momentum values, and thus, its contribution to the momentum density can be neglected, since it gives contribution only to the $1/p^{12}$ term.

Next, we can average the momentum density over the momentum orientations, \emph{i.e.,}
integrate out the angular dependence of the $\bp$ vector and divide by $4 \pi$, that reads for the second term in the square bracket of \rref{eq:rhopvec} as
\begin{align}
    \frac{1}{4\pi} 
    \int_0^{2 \pi} \mbox{d} \varphi 
    \int_{-1}^1 \mbox{d}(\cos{\vartheta}) \,  \cos \left[%
        \bos{p} \left( {\bf R}_{A}-{\bf R}_B\right)
        \right]   
    =
    \frac{1}{2}  
    \int_{-1}^1 \mbox{d}c\,  \cos \left( p R_{AB} c\right) 
    =
    \frac{ %
      1
    }{%
      p R_{AB}
    }
    \sin \left( p R_{AB}\right) \; ,
\end{align}
and thereby, we obtain the spherically averaged momentum density, 
\begin{align}
\text{for } p>p_\Lambda:\quad 
   \left \langle \bar{\rho} \right \rangle_{\vartheta,\varphi}(p)
   &=
   \frac{1}{4\pi} 
    \int_0^{2 \pi} \mbox{d} \varphi 
    \int_{-1}^1 \mbox{d} 
      (\cos{\vartheta}) \, \bar{\rho}(\bos{p}) \nonumber \\ %, 
    &= 
    \frac{2}{ \pi p^8} 
    \Bigg[ %
    4 \sum_{A=1}^{N_\text{nucl}} Z_A^2 \, \rho(\bos{R}_A) 
    + 
    4\sum_{A=1}^{N_\text{nucl}} 
    \sum_{B \ne A}^{N_\text{nucl}}
      Z_A Z_B 
    \frac{%
      \sin\left( p R_{AB}\right)
    }{%
      p R_{AB}
    } 
    \Gamma(\bos{R}_A,\bos{R}_B) 
    + 
    \eta(\bos{0}) 
    \Bigg] 
    %\nonumber \\ 
    %& \hspace{12cm} 
    + \mathcal{\bos{O}}\left( p^{-10}\right) \ . \label{finmomdens} 
\end{align}

\subsection{Asymptotic tail of \texorpdfstring{$\itfun{p^k}(p)$}{} and its contribution to \texorpdfstring{$\llangle \Psi | p_1^k |\Psi\rrangle$}{}}
  Using the derived large-momentum, asymptotic tail of the momentum density, \rref{finmomdens},
  we can calculate its contribution to the asymptotic tail of $\itfun{p^k}(p)$, Eqs.~(\ref{pkexpval})--(\ref{eq:pkshl}),
  \begin{align}
      \text{for } p>p_\Lambda:&  \nonumber  \\
      \titfun{p^k}(p) 
      &= 
     \frac{ 4\pi}{N} p^{k+2} 
     \left \langle \bar{\rho} \right \rangle_{\vartheta,\varphi}(p) \nonumber \\
      &=
      \frac{8}{ p^{6-k} {N}} \Bigg[ %
        4 \sum_{A=1}^{{N_\mathrm{nucl}}} Z_A^2 \, \rho(\bos{R}_A) 
        + 
        4 \sum_{A=1}^{{N_\mathrm{nucl}}}\sum_{B \ne A}^{{N_\mathrm{nucl}}} 
          Z_A Z_B 
          \frac{%
            \sin\left( p R_{AB}\right)
          }{%
            p R_{AB}
          } 
          \Gamma(\bos{R}_A,\bos{R}_B) 
          +  \eta(\bos{0}) \Bigg] \nonumber \\
      & \hspace{9cm}+  \sum_{j=1}^m \frac{A_j}{p^{7-k+j}} +
  \mathcal{O}\left( p^{-8+k-m}\right) \ , \label{nonintIpkansatz}
  \end{align}
  where the $A_j$ coefficients are determined by fitting and $m$ is chosen to fix the number of additional terms considered in the expansion.
  In our calculations the typical value for $m$ was between 4 and 7.  
  
  Using these expressions, 
  the contribution from the large-momentum tail to 
  $\llangle \Psi | p_1^k |\Psi\rrangle$ in Eq.~(\ref{eq:pkshl}), can be calculated. 
In this paper, we focus on the $k=2$ and $k=4$ cases, for which the final expression is 
\begin{align}
  \int_{p_\Lambda}^\infty \dd p\ \titfun{p^k}(p) 
    &= 
    \frac{8}{ (5-k) p_\Lambda^{5-k} N} \Bigg[% 
    4 \sum_{A=1}^{N_\text{nucl}} Z_A^2 \, \rho(\bos{R}_A) + \eta(\bos{0}) \Bigg]  \nonumber \\
     & \hspace{1cm} 
     + \frac{32}{N} \sum_{A=1}^{N_\text{nucl}} \sum_{B=1}^{N_\text{nucl}} 
       Z_A Z_B \Gamma(\bos{R}_A,\bos{R}_B) 
       %G_k(p_\Lambda,\left| \bos{R}_A -\bos{R}_B \right| )\nonumber \\
       G_k(p_\Lambda,R_{AB})\nonumber \\
     & \hspace{1cm} 
       + \sum_{j=0}^m \frac{A_j}{(6-k+j)p_\Lambda^{6-k+j}N } +
  \mathcal{O}\left( p_\Lambda^{-7+k-m}\right) \ ,  \label{Ipkansatz}
\end{align}
with
\begin{align}
     G_2(p_\Lambda,R) &=  \frac{\cos\left(R p_\Lambda \right)}{2 p_\Lambda} + \frac{ \sin \left( R p_\Lambda \right)}{2 p_\Lambda^2 R} - 
    \frac{R \left[ \pi - 2 \mathrm{Si} \left(R p_\Lambda  \right) \right])}{4} \quad \text{for}\quad k=2 ,\\
     G_4(p_\Lambda,R) &= \frac{R^2 p_\Lambda^2-2}{24 p_\Lambda^3} \cos \left(R p_\Lambda \right) -\frac{R^2 p_\Lambda^2-6}{24 R p_\Lambda^4} \sin \left(R p_\Lambda \right) +  \frac{R^3   \left[ \pi - 2 \mathrm{Si} \left(R p_\Lambda  \right) \right])}{48} \   \quad \text{for}\quad k=4  \ ,
\end{align}
where $\mathrm{Si}(x)$ is the sine integral function \cite{f.w.j.olverNISTDigitalLibrary2021d}.

In the  numerical calculations, $\rho(\bos{R}_A)$ and $\eta(\bos{0})$ are determined by using the method described in Sec.~\ref{sec:ITCD},
\begin{align}
    \rho(\bos{R}_A) &= \sum_{i=1}^N  \left \langle \Psi \left| \delta\left( \bos{r}_i -\bos{R}_A\right)\right | \Psi  \right \rangle \ , \\
    \eta(\bos{0}) &= \sum_{i=1}^N  \sum_{j>i}^N  \left \langle \Psi \left| \delta \left( \bos{r}_i -\bos{r}_j \right) \right| \Psi \right \rangle \ .
\end{align}
The quantity $\Gamma(\bos{R}_A,\bos{R}_B)$ is an element of the density matrix, for which the cusp condition is also known \cite{clintonCuspConditionConstraint1972,davidson_reduced_2014}, but it is handled as a fitting parameter in the present work.

\begin{figure}
    \centering
    \includegraphics[scale=1]{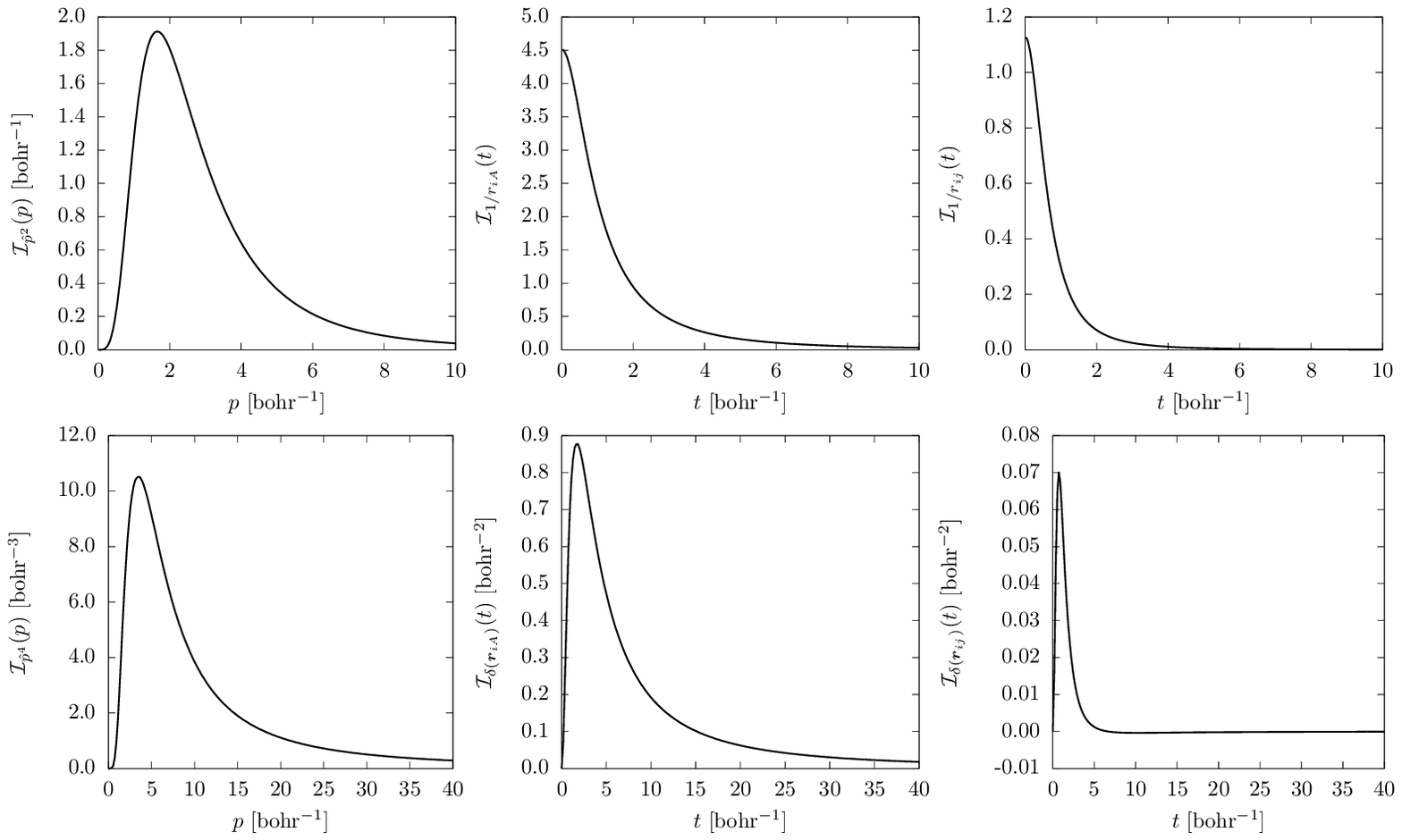}
    \caption{%
       IT function profile for various operators, 
       $\itfun{\hat{p}^2}$, 
       $\itfun{1/r_{iA}}$, 
       $\itfun{1/r_{ij}}$,
       $\itfun{\hat{p}^4}$,
       $\itfun{\delta(\br_{iA})}$, and 
       $\itfun{\delta(\br_{ij})}$ 
       for the example of the ground electronic state of the helium atom. 
    }
    \label{fig:itcurvesHe}
\end{figure}
\begin{figure}
    \centering
    \includegraphics[scale=1]{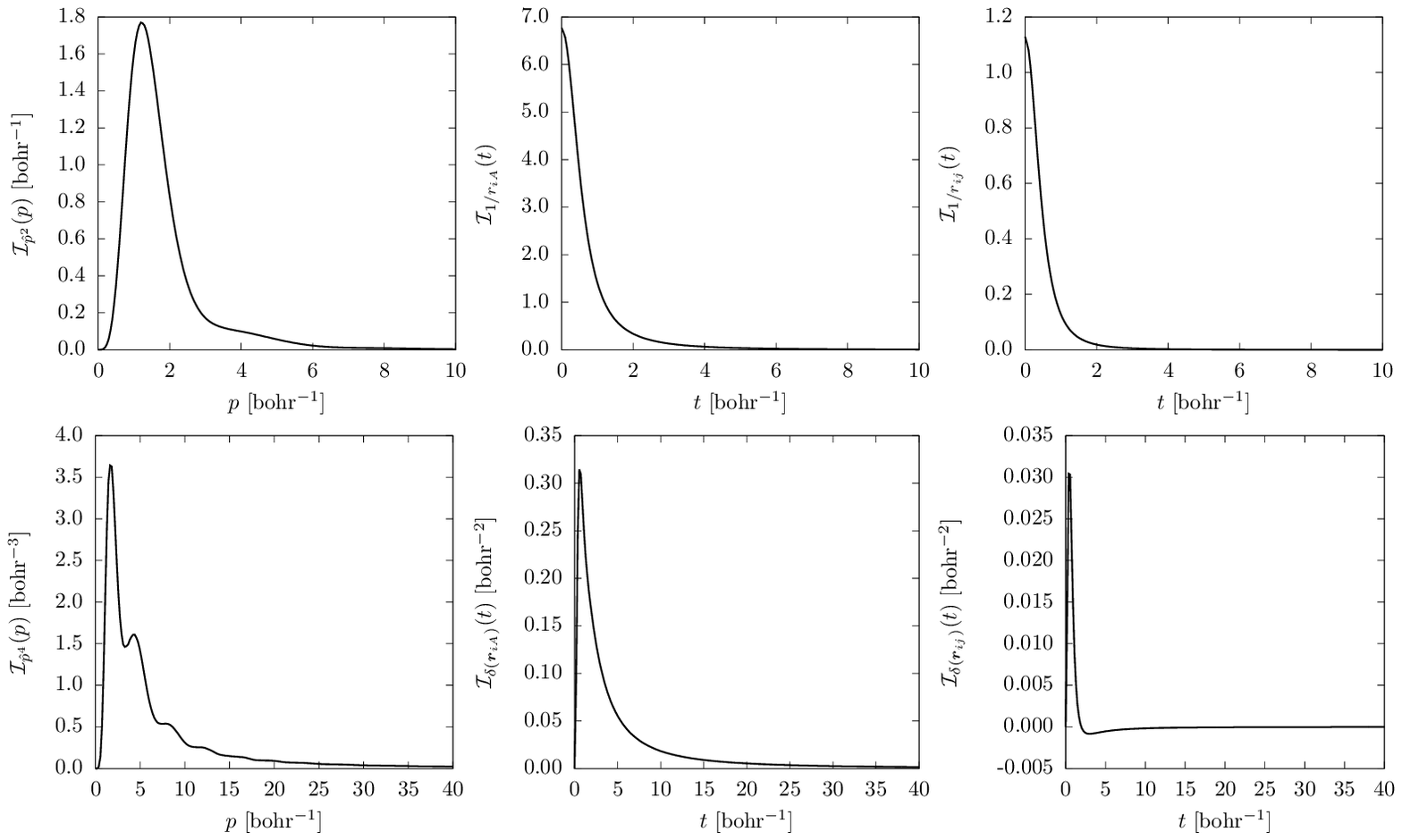}
    \caption{%
       IT function profile for various operators, 
       $\itfun{\hat{p}^2}$, 
       $\itfun{1/r_{iA}}$, 
       $\itfun{1/r_{ij}}$,
       $\itfun{\hat{p}^4}$,
       $\itfun{\delta(\br_{iA})}$, and 
       $\itfun{\delta(\br_{ij})}$ 
       for the example of the ground electronic state of the H$_3^+$ with protons (p) clamped at an equilateral triangular configuration with $R_\text{pp}=1.65$~bohr. }
    \label{fig:itcurvesH3p}
\end{figure}

\section{Computational details\label{sec:implem}}
\noindent
The integral transformed functions  
$\itfun{\hat{p}^2}$, $\itfun{\hat{p}^4}$, $\itfun{1/r_{iA}}$, $\itfun{1/r_{ij}}$, $\itfun{\delta(\br_{iA})}$, and $\itfun{\delta(\br_{ij})}$ are shown in Figures~\ref{fig:itcurvesHe} and \ref{fig:itcurvesH3p} for the example of the ground electronic state of the helium atom (He) and the trihydrogen cation (H$_3^+$) with protons (p) clamped at an equilateral triangular configuration with $R_\text{pp}=1.65$~bohr. 

Up to a certain $\xi_\Lambda$ value  ($\xi_\Lambda=p_\Lambda$ for momentum operators, and $\xi_\Lambda=t_\Lambda$ for Coulombic operators), we calculate the short-range integral analytically for $\itfun{\hat{p}^2}$, $\itfun{1/r_{iA}}$,  $\itfun{1/r_{ij}}$, and by quadrature for $\itfun{\hat{p}^4}$, $\itfun{\delta(\br_{iA})}$, and 
$\itfun{\delta(\br_{ij})}$ (for more details see Appendix \ref{sec:Gint}).  For the long-range part, it is necessary to determine the accurate value of $\rho(\bos{0})$ and $\eta(\bos{0})$, which is calculated by an iterative procedure using \rrefsa{eq:asymdelta},\rrefsb{eq:ITdelta}, \rrefsb{intIdeltarij}, and \rrefsb{eq:ITdeltarij}.
Then, the long-range part is obtained by fitting the asymptotic tail to data points using \rrefsa{eq:asymcoulomb}, \rrefsb{eq:asymcoulombrij}, and \rrefsb{nonintIpkansatz} that is followed by  
the analytic integration of the asymptotic tail,  \rrefsa{intIri}, \rrefsb{intIrij}, and \rrefsb{Ipkansatz}, using the fitted parameters. 

It is critical to choose an optimal $\xi_\Lambda$ value and a good interval for the data used for the fitting of the long-range analytic expression. We have selected these parameters based on the inspection of the integrand evaluated with the approximate wave function (Figs.~\ref{fig:itcurvesHe} and \ref{fig:itcurvesH3p}). 
Close to the origin, the asymptotic expansion fails, but the ECG basis describes well the non-analytic correlation effects in this range. The parameter $\xi_\Lambda$ must be large enough to ensure that the function $\itfun{F(\bos{r}_i)}(\xi_\Lambda)$  can be approximated accurately with the asymptotic expansion. At the same time, it must be small enough to eliminate the major numerical uncertainties from the finite basis expansion. For the spherically symmetric ground state of the helium atom (Fig.~\ref{fig:itcurvesHe}), $\itfun{}(\xi)$ is simple, it decreases monotonically to zero  after an initial peak. The asymptotic part can be `easily' identified and fitted to the asymptotic series. The H$_3^+$ molecular ion (Fig.~\ref{fig:itcurvesH3p}) is a more `complex' system, with more complicated correlation effects, and thus, we need to choose a larger $\xi_\Lambda$ value to reach the asymptotic regime (which also implies the use of a larger basis set). Further details about the accuracy of the matrix elements depending on the selection of the $\xi_\Lambda$ value can be found in Sec. \ref{sec:nonrel}. 

According to Secs.~\ref{sec:longrange} and \ref{sec:asympdens}, the long-range part of the function $\itfun{}(\xi)$ decays polynomially due to the cusp in the exact wave function that is approximated in the computations. At the same time, 
we may observe in Figure~\ref{fig:fit} that the approximate $\itfun{p^k}(p)\ (k=2,4)$ function, 
corresponding to a finite ECG basis set, 
has artificial oscillations in momentum space and some non-negligible deviations in $t$-space. 
If the full integral is computed by direct integration, the oscillations approximately cancel in the integral, and this explains the practical observation that accurate results can be obtained even with ECGs that fail to satisfy analytic properties of the exact wave function.
We aim to obtain more accurate integral values by replacing the oscillatory asymptotic tail with the mathematically correct decaying form corresponding to the cusp.

In practice, the numerical accuracy of the computations is affected by the grids used for the numerical integration (if analytic integration is not possible over the finite, short-range interval) and for the fitting procedure. 
Our computational strategies are explained in the following sections for the physical operators considered in this work.

\section{Perturbative relativistic correction for \texorpdfstring{H$_3^+$}{}  at equilibrium \label{sec:rel}}
For the spectroscopic characterization of compounds of light elements,
the leading-order relativistic correction has been traditionally calculated as the expectation value of the Breit--Pauli Hamiltonian with the non-relativistic wave function. 
The Breit--Pauli Hamiltonian is the the leading-order Foldy--Wouthuysen perturbation theory (FWPT) term of the Dirac--Coulomb--Breit  Hamiltonian \cite{dyall_introduction_2007,reiher_relativistic_2015,cencek_accurate_1996}. 
The singular operators that are difficult to evaluate in a Gaussian basis appear already for the 
the leading-order FWPT of the Dirac--Coulomb operator that reads for the two electrons of H$_3^+$ with fixed protons ($N=2$ and $N_\text{nucl}=3$) as 
\begin{align}
    \hat{H}^\text{FW} &= 
    \hat{H}_\mathrm{nonrel} + \Delta\hat{H}^\text{FW}
    \label{eq:FW}\\
    c^2\Delta\hat{H}_\mathrm{DC}^\mathrm{FW} &=  
       \underbrace{%
       -\frac{1}{8}\sum_{i=1}^N \nabla_i^4
       }_{\text{mass-velocity}}
       \underbrace{%       
       + \frac{\pi}{2}\sum_{i=1}^N\sum_{A=1}^{N_\mathrm{nucl}} Z_A \delta(\br_{iA}) 
       }_{\text{Darwin I}}       
       \underbrace{%           
       - \pi \sum_{i=1}^N\sum_{j>i}^N \delta(\br_{ij})
       }_{\text{Darwin II}}
      \ , \label{eq:FWDC} \\
    c^2\Delta\hat{H}_\mathrm{DCB}^\mathrm{FW} &=  
    c^2\Delta\hat{H}_\mathrm{DC}^\mathrm{FW} 
       \underbrace{%
       + 2\pi \sum_{i=1}^N\sum_{j>i}^N \delta(\bos{r}_{ij})
       }_{\text{spin-spin}}
       \underbrace{%
       - 
       \sum_{i=1}^N\sum_{j>i}^N
       \frac{1}{2r_{ij}}
       \left(%
         \bos{p}_i\bos{p}_j
         +
         \frac{\bos{r}_{ij}(\bos{r}_{ij}\bos{p}_i)\bos{p}_j}{r_{ij}^2}
       \right)
       }_{\text{orbit-orbit}}
       \nonumber \\
    &=
       -\frac{1}{8}\sum_{i=1}^N \nabla_i^4 
       + 
       \frac{\pi}{2}\sum_{i=1}^N\sum_{A=1}^{N_\mathrm{nucl}} Z_A \delta(\br_{iA}) 
       +\pi \sum_{i=1}^N\sum_{j>i}^N \delta(\br_{ij})
       -
       \sum_{i=1}^N\sum_{j>i}^N 
       \frac{1}{2r_{ij}}
       \left(%
         \bos{p}_i\bos{p}_j
         +
         \frac{\bos{r}_{ij}(\bos{r}_{ij}\bos{p}_i)\bos{p}_j}{r_{ij}^2}
       \right)
      \ , \label{eq:FWDCB}
\end{align}
for singlet states.
$\hat{H}_\mathrm{nonrel}$ is the non-relativistic Hamiltonian and $\Delta\hat{H}_\mathrm{DC}^\text{FW}$ 
and
$\Delta\hat{H}_\mathrm{DCB}^\text{FW}$
correspond to the leading-order correction to the non-relativistic energy of the Foldy--Wouthuysen (FW) transformed Dirac--Coulomb (DC) and Dirac--Coulomb--Breit (DCB) Hamiltonians.

We have calculated the expectation value of the mass-velocity and the Darwin terms with the non-relativistic wave function both by direct integration and by the integral transform (IT) technique. IT technique for the example of the simplest polyatomic molecule, H$_3^+$ near its equilibrium structure (Table~\ref{tab:ptterms}). 
In Table~\ref{tab:ptterms}, we also show the (non-singular) orbit-orbit term [last term in \rref{eq:FWDCB}] by direct integration.

Regarding the computational parameters, the $\xi_\Lambda=100$~bohr$^{-1}$ threshold value was appropriate also in this system, similarly to the He and H$_2$ computations reported in Ref.~\cite{pachucki_acceleration_2005}. 
The short-range integrals 
were calculated by quadrature. 
For the Dirac-delta terms, the numerical integration was carried out over three subintervals, $[0,1]$ bohr$^{-1}$, $[1,10]$ bohr$^{-1}$, and $[10,100]$ bohr$^{-1}$, using 
25, 35, and 35 Gauss--Legendre quadrature points. For the mass-velocity term, 
we have checked the convergence of the integral value over the $[10,100]$ bohr$^{-1}$ interval using 50, 70, and 100 number of points.
The value of the integrand at each grid point is obtained 
by direct evaluation of the finite basis ECG integral (Appendix~\ref{sec:Gint}). This setup was sufficient for a parts-per-billion (ppb) convergence of the short-range integral value.

For fitting the long-range part of the integrand, we have considered the $p>60$~bohr$^{-1}$ tail that is beyond the range dominated by non-trivial correlation effects (Fig.~\ref{fig:itcurvesH3p}).
We have carried out the fitting of the asymptotic tail by including additional grid points from the [100,390] bohr$^{-1}$ interval with 10 bohr$^{-1}$ spacing.
In each fit, six parameters were included, and the squared sum of residuals was on the order of $10^{-20}$ (a.u.) for $\delta(\br_{ij})$ and $\delta(\br_{iA})$ 
and $10^{-10}$ (a.u.) for the $p_1^4+p_2^4$ term.
Table~\ref{tab:ptterms} collects the terms appearing in the relativistic corrections obtained with direct integration and by the IT technique that reduces the relative error of the expectation value by ca.~2 orders of magnitude.

In Table \ref{tab:energ}, 
the leading-order FW-DC and FW-DCB energy
is compared with the no-pair variational energy of the corresponding (DC and DCB) operators \cite{jeszenszki_all-order_2021}. {In the perturbative DC energy, we observe an error cancellation for the singular terms, hence, the direct and the IT FW-DC energies differ only on the order of 1~nE$_\text{h}$. For the FW-DCB energy, due to the spin-spin contribution, \rref{eq:FWDCB}, there is a 15~nE$_\text{h}$ deviation between the direct and the IT results. 
For comparison, we also show the variational DC(B) energies \cite{jeszenszki_all-order_2021} that 
are not affected by the slow convergence problem of the singular operators. 
A detailed comparison of the variational and the perturbative FW energies will be provided in future work.
}

\begin{table}
  \caption{%
    Expectation value of operator terms in the leading-order Foldy--Wouthuysen perturbative relativistic operators (in atomic units)
    computed by direct integration (`Direct') and by the integral transformation technique (`IT')
    for the ground electronic state of H$_3^+$ with protons (p) clamped at an equilateral triangular configuration with $R_\text{pp}=1.65$~bohr. 
    The basis set size corresponds to the use of $D_{3\text{h}}$
    point-group symmetry in the computations.    
    \label{tab:ptterms}
  }
  \begin{tabular}{@{}l@{\ \ \ } l@{\ \ \ }lc l@{\ \ \ }lc l@{\ \ \ }l l@{\ \ \ }l @{}}
    \hline\hline\\[-0.35cm]
    &	
    \multicolumn{2}{c}{$\langle \nabla_1^4 + \nabla_2^4\rangle$} & &
    \multicolumn{2}{c}{$\sum_{i=1}^2\sum_{A=1}^3 Z_A \left \langle \delta \left( \br_i - \bR_A \right)\right \rangle$} & &
    \multicolumn{2}{c}{$\left \langle \delta \left( \br_1 - \br_2 \right)\right \rangle$} 
    &&
    \multicolumn{1}{c}{\raisebox{-0.15cm}{Orbit-orbit term}}
    \\
    \cline{2-3} \cline{5-6} \cline{8-9}\\[-0.3cm]
    \multicolumn{1}{l}{$\nb$} &	
    \multicolumn{1}{c}{Direct} &
    \multicolumn{1}{c}{IT} &~~~&  
    \multicolumn{1}{c}{Direct} &
    \multicolumn{1}{c}{IT} &~~~&  
    \multicolumn{1}{c}{Direct} &
    \multicolumn{1}{c}{IT} && \\
    \cline{1-11}\\[-0.3cm]
     150 & 15.428 820 & {15.467 265} 
     && 1.086 786 273 & 1.089 641 891 && 0.018 430 054 & 0.018 340 790 && $-$0.057 219 009  \\
     200 & 15.446 739 & {15.467 346} 
     && 1.088 110 465 & 1.089 651 086 && 0.018 407 593 & 0.018 336 611 && $-$0.057 218 310  \\
     300 & 15.455 982 & {15.467 351} 
     && 1.088 821 792 & 1.089 654 339 && 0.018 368 291 & 0.018 335 079 && $-$0.057 217 628  \\
     400 & 15.456 244 & {15.467 368} 
     && 1.088 836 952 & 1.089 654 512 && 0.018 360 864 & 0.018 334 828 && $-$0.057 217 548   \\
     500 & 15.456 360 & {15.467 395} 
     && 1.088 843 368 & 1.089 654 577 && 0.018 358 011 & 0.018 334 777 && $-$0.057 217 524   \\
     600 & 15.456 386 & {15.467 395}
     && 1.088 845 002 & 1.089 654 597 && 0.018 357 565 & 0.018 334 773 && $-$0.057 217 520  \\
    \hline\hline
  \end{tabular}
\end{table}

\begin{table}
  \caption{%
    Non-relativistic, perturbative ($E^\text{FW}_\text{DC}$ and $E^\text{FW}_\text{DCB}$) and no-pair variational ($E^\text{npV}_\text{DC}$ and $E^\text{npV}_\text{DCB}$) relativistic electronic energies, in $\Eh$, 
    for the ground electronic state of H$_3^+$ with protons (p) clamped at an equilateral triangular configuration with $R_\text{pp}=1.65$~bohr (see also caption to Table~\ref{tab:ptterms}). 
    We used the speed of light $c=\alpha^{-1}a_0\Eh/\hbar$ with 
    $\alpha^{-1}=137.$035 999 084 \cite{Codata2018Recommended}.
    \label{tab:energ} 
  }
  \begin{tabular}{@{}lcccc @{\ \ \ \ } c c c c @{}}
    \hline\hline\\[-0.35cm] &
    \multicolumn{1}{c}{$\nb$} &
     \multicolumn{1}{c}{$E_\mathrm{nonrel}$} &
    \multicolumn{1}{c}{$E_\text{DC}^\mathrm{FW}(\text{Direct})^\text{a}$}  & 
    \multicolumn{1}{c}{$E_\text{DC}^\mathrm{FW}(\text{IT})^\text{a}$} & 
    $E^\text{npV}_\text{DC}$ \cite{jeszenszki_all-order_2021} &
    \multicolumn{1}{c}{$E_\text{DCB}^\mathrm{FW}(\text{Direct})^\text{b}$} &
    \multicolumn{1}{c}{$E_\text{DCB}^\mathrm{FW}(\text{IT})^\text{b}$} & 
    \multicolumn{1}{c}{$E_\text{DCB}^\mathrm{npV}$} \cite{jeszenszki_all-order_2021} \\
    \cline{1-9}\\[-0.3cm]
     & 150 &  $-$1.343 835 557  & $-$1.343 850 435 & $-$1.343 850 437
     & $-$1.343 850 149  & $-$1.343 847 315 & $-$1.343 847 347
     & $−$1.343 847 343 \\
     & 200 &  $-$1.343 835 606  & $-$1.343 850 488 & $-$1.343 850 485
     & $-$1.343 850 507  & $-$1.343 847 376 & $-$1.343 847 396
     & $−$1.343 847 404 \\
     & 300 &  $-$1.343 835 623  & $-$1.343 850 501 & $-$1.343 850 501
     & $-$1.343 850 524  & $-$1.343 847 402 & $-$1.343 847 413
     & $−$1.343 847 462 \\
     & 400 &  $-$1.343 835 624 & $-$1.343 850 502 & $-$1.343 850 502
     & $-$1.343 850 526   & $−$1.343 847 405 & $−$1.343 847 415
     & $−$1.343 847 484 \\
     & 500 &  $-$1.343 835 625 & $-$1.343 850 502 & $-$1.343 850 503
     & $-$1.343 850 527   & $−$1.343 847 406 & $−$1.343 847 416
     & $−$1.343 847 496 \\
     & 600 &  $-$1.343 835 625 & $-$1.343 850 502 & $-$1.343 850 503
     & $-$1.343 850 527   & $-$1.343 847 406 & $-$1.343 847 416
     & $−$1.343 847 498 \\
    \hline\hline
  \end{tabular}
  \begin{flushleft}
    $^\text{a}$~Expectation value of $\hat{H}^{\text{FW}}_\text{DC}$, Eqs.~(\ref{eq:FW}) and (\ref{eq:FWDC}), with the non-relativistic wave function. \\
    $^\text{b}$~Expectation value of $\hat{H}^{\text{FW}}_\text{DCB}$, Eqs.~(\ref{eq:FW}) and (\ref{eq:FWDCB}), with the non-relativistic wave function.    
\end{flushleft}  
\end{table}

\section{An attempt to improve the non-relativistic energy with the integral transformation technique \label{sec:nonrel}}
According to Sections~\ref{sec:ITCD} and \ref{sec:FTkin}, the integral transformation technique can be used also for the expectation values of the non-relativistic operators, $\langle \hat{p}^2 \rangle$,  $\langle 1/ r_{iA} \rangle$, and  $\langle 1/r_{ij} \rangle$. Although these expectation values converge (much) faster than the expectation value of the singular operators appearing in the relativistic corrections, pinpointing their precise value would be useful to have an (even) better estimate of the complete basis limit. In this section, we report observations of some exploratory work for the $\langle \hat{p}^2 \rangle$ operator for the ground state of the helium atom. 

Thanks to the simplicity of the $\itfun{\hat{p}^2}$,  $\itfun{1/\hat{r}_{iA}}$, and $\itfun{1/\hat{r}_{ij}}$ integrands, 
the short-range integrals can be obtained in an analytic form (Appendix~\ref{sec:Gint}).
To fit the $\titfun{\hat{O}}(\xi)$ asymptotic part, an equidistant grid was used. The start of the fitting interval was determined based on inspection of the integrand functions (Fig.~\ref{fig:itcurvesHe}). 
On the one hand, we wanted to choose a large $\xi$ value to avoid fitting to non-trivial correlation features. On the other hand, we wanted to choose a small $\xi$ value to reduce the finite basis error of the ECG basis set.
A short summary about the calculation of the necessary $\rho(\bos{0})$ and $\eta(\bos{0})$ values is provided in Appendix~\ref{sec:rhoeta}.
Figure~\ref{fig:fit} shows the relative difference 
of $\itfun{\hat{O}}(\xi)$ represented by the finite basis expansion 
and by the analytically known leading-order asymptotic part, 
$\titfun{p^k}^0(p)= N^{-1} p^{-2-k} 128 \pi \rho(\bos{0})$,  
$\titfun{1/r_{iA}}^0(t)= N^{-1} t^{-3}(2\pi\rho(\bos{0})-16\sqrt{\pi} \rho(\bos{0}) t^{-1})$, 
$\titfun{1/r_{ij}}^0(t)=[N(N-1)]^{-1} t^{-3} 2(\pi\eta(\bos{0})+ 2\sqrt{\pi} \eta(\bos{0}) t^{-1})$. 

For larger (but not too large, \emph{i.e.,} for which the finite basis representation can be trusted) $\xi$ values, the relative difference is determined by the contributions beyond the analytic, leading-order terms. The deviation from zero in the asymptotic limit indicates numerical errors, which originate from the finite basis-set approximation. 

For the Coulomb terms, $\titfun{1/r_{ij}}$ and $\titfun{1/r_{iA}}$ (Figs.~\ref{fig:fit}c, d), this numerical error is monotonic and has non-negligible values beyond some $t$ value, but by increasing the basis set size, this critical $t$ threshold is shifted towards larger values. 

\begin{figure}
   \includegraphics[scale=1]{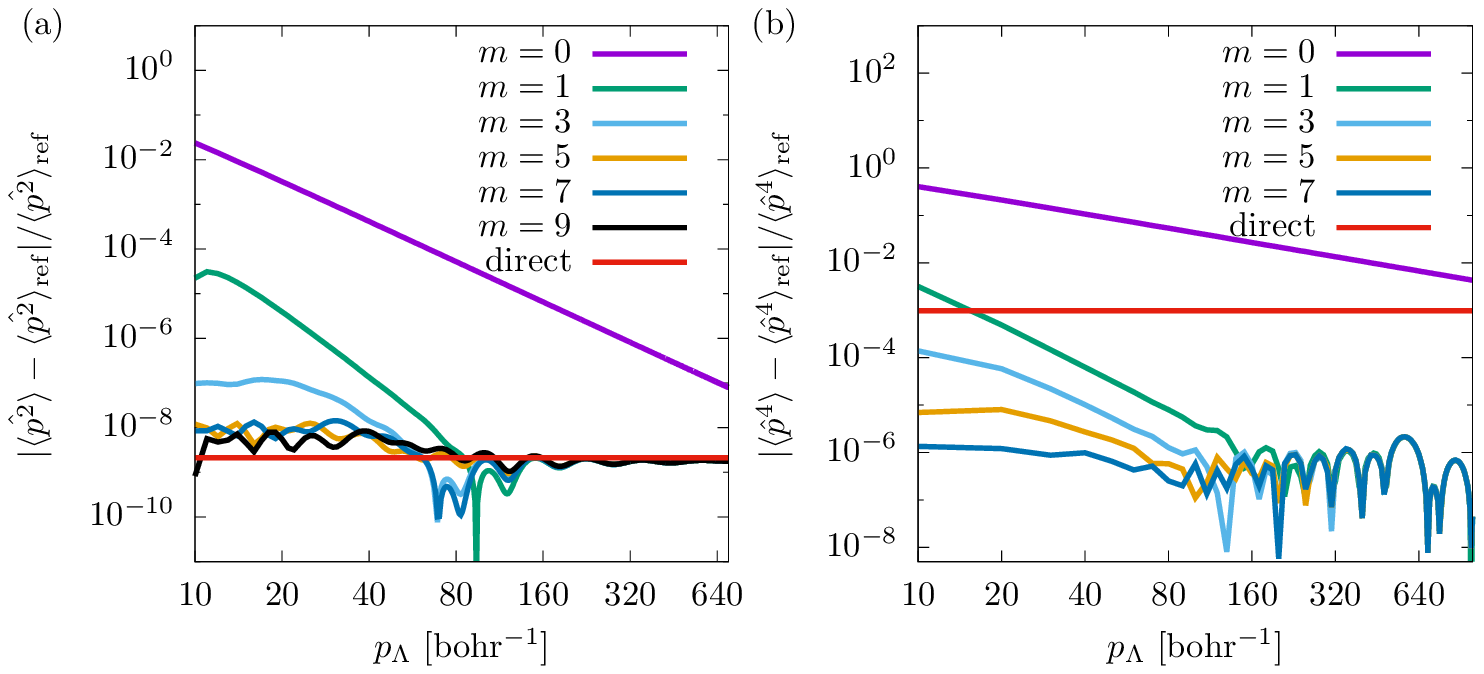}
  \caption{%
Relative error of $\llangle\hat{p}^2\rrangle$  and $\llangle\hat{p}^4\rrangle$ for helium, for various $p_\Lambda$ thresholds and $m$ terms in fitting function \rref{Ipkansatz}. The result `direct'
was obtained by direct integration with ECGs. 
The reference values are
$\langle \hat{p}^2 \rangle_\mathrm{ref}=2.903\ 724\ 377\ 034\ 119\ 5$~bohr$^{-2}$ \cite{drake_high_2006}  
(using the virial theorem $\langle \hat{T} \rangle =-E$), 
and 
$\langle \hat{p}^4 \rangle_\mathrm{ref}=108.176\ 134\ 4(8)$~bohr$^{-4}$ \cite{pachucki_acceleration_2005}. \label{fig:Terror}
  }
\end{figure}

Regarding the $\hat{p}^k$ operators (Figs.~\ref{fig:fit}a, b), the Fourier transformation results in oscillations that can be observed for the finite-basis representation of $\itfun{\hat{p}^k}$ over the large momentum range. The oscillation amplitude decreases upon increasing the basis set size.
Figure~\ref{fig:Terror} shows the comparison of the direct and the IT integration procedures 
for $\hat{p}^2$ and $\hat{p}^4$.
The computational details for $\hat{p}^4$ can be found in the Sec.~\ref{sec:implem}. 
Regarding $\hat{p}^2$, the fit is performed over the $[10,90]$~bohr$^{-1}$
interval using 1600 equidistant points. Depending on the number of the fitting parameters the squared sum of the residuals varied between $10^{-11}$ and $10^{-17}$ (a.u.).

The effect of the choice of the $p_\Lambda$ threshold value, 
which separates the short- and the long-range intervals, 
and the number of the fitted parameters in the long-range part 
is shown in Figure~\ref{fig:Terror}.
For both $\hat{p}^2$ and $\hat{p}^4$, the larger the number of the fitted parameters, the better results are observed, especially for smaller $p_\Lambda$ values. By increasing $p_\Lambda$, all curves are close, since the high-order inverse momentum terms have a numerically negligible contribution in this regime. For $p_\Lambda\rightarrow\infty$, the contribution from the integral transformation goes to zero, and the direct integration result is recovered. 
It is also necessary to note that although we achieve a better relative accuracy for $\llangle \hat{p}^2 \rrangle$ than for $\llangle \hat{p}^4 \rrangle$,  the improvement of 
$\llangle \hat{p}^2 \rrangle$ (IT) over $\langle \hat{p}^2 \rangle$ (`direct') is modest.
This observation can be contrasted with the two orders of magnitude improvement
of 
$\llangle \hat{p}^4 \rrangle$ (IT) over $\langle \hat{p}^4 \rangle$ (`direct')
that appears to be a robust feature with respect to the choice of $p_\Lambda$ and the fitting details.
For $\hat{p}^2$, the `optimal' interval for $p_\Lambda$ and the fitting details should be very carefully chosen to observe any improvement.

\begin{figure}
  \includegraphics[scale=1]{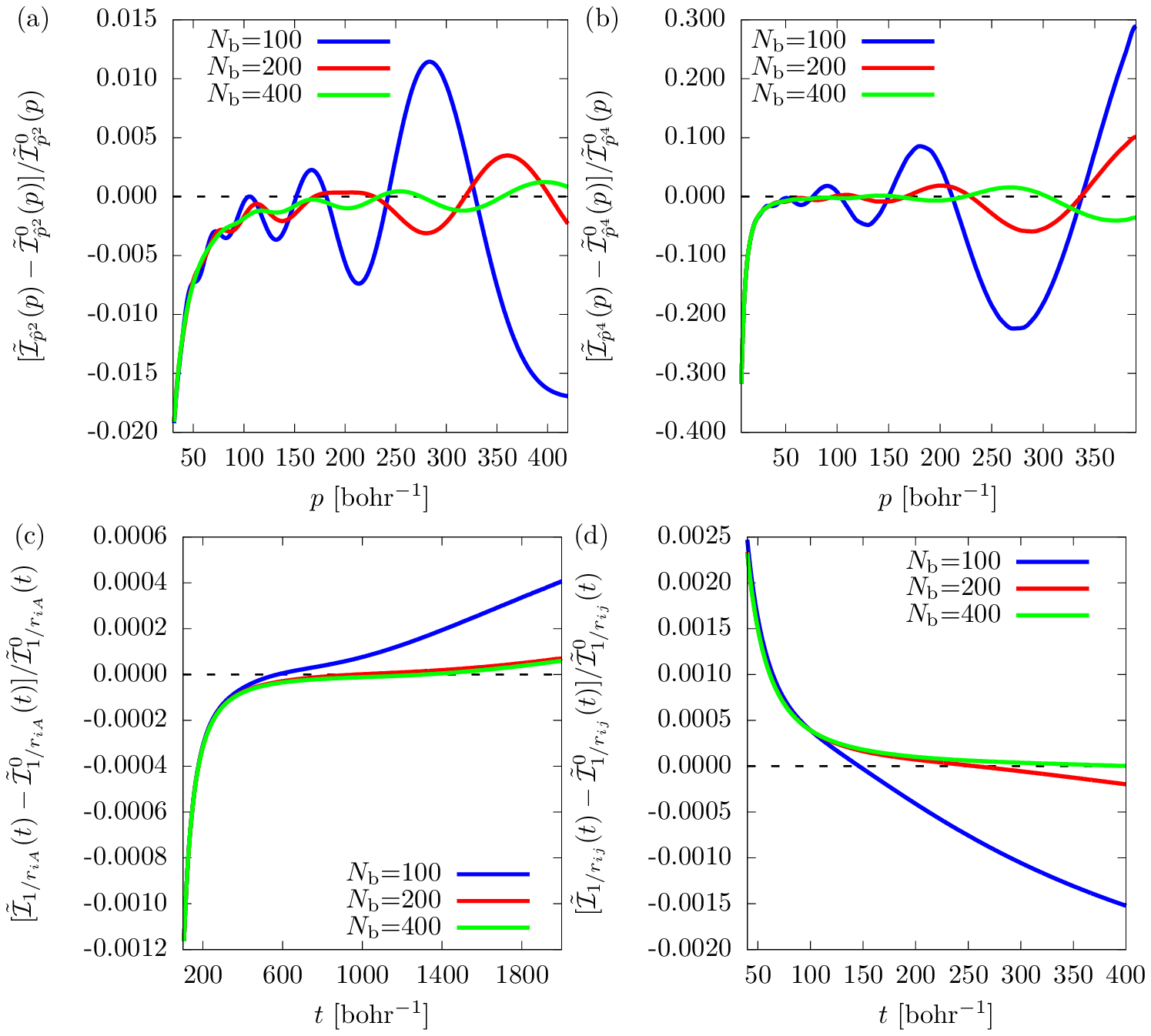}
  \caption{%
    Relative difference in the asymptotic tail of 
    the numerically calculated functions $\itfun{p^2}(p)$, $\itfun{p^4}(t)$, $\itfun{1/r_{iA}}(t)$, $\itfun{1/r_{ij}}(t)$, 
    and 
    the analytic leading-order expressions of the asymptotic tail, $\titfun{p^k}^0(p)= N^{-1}p^{-2-k} 128 \pi \rho( \bos{0})$, 
    $\titfun{1/r_{iA}}^0(t)=N^{-1}t^{-3}(2\pi\rho(\bos{0})-16\sqrt{\pi} \rho(\bos{0}) t^{-1})$, 
    $\titfun{1/r_{ij}}^0(t)=[N(N-1)]^{-1}t^{-3}
    2(\pi\eta(\bos{0})+2\sqrt{\pi} \eta(\bos{0}) t^{-1})$  
    for the example of the ground state of the helium atom with an increasing number of ECG basis functions ($N_\text{b}$). 
  \label{fig:fit}
  }
\end{figure}

\clearpage
\section{Summary and conclusion\label{sec:summary}}
\noindent Explicitly correlated Gaussian functions (ECGs) are often used in atomic and molecular computations, since they incorporate particle-particle correlation and they have analytic integrals for most physically relevant operators.
In spite of their advantages, they also have some drawbacks. 
They fail to describe correctly the particle coalescence points and the asymptotic tail of the exact non-relativistic wave function of Coulomb-interacting point-like particles.
This paper was devoted to the study of a possible correction scheme for coalescence properties during computations with Gaussian-type functions.

For this purpose, we have presented the detailed theoretical background of the integral transformation (IT) technique originally proposed by Pachucki, Cencek, and Komasa \cite{pachucki_acceleration_2005} to enhance the convergence of singular operators appearing in perturbative relativistic corrections. 
The core idea of the IT technique is to rewrite expectation values of physical quantities with an approximate wave function into a form, for which the cusp effect---characteristic for short ranges in coordinate space---appears in the asymptotic tail of the integrand in an `inverse space' ($\xi$). For momentum-type operators, this transformation is the Fourier transformation and the inverse space is momentum space ($\xi=p$). For Coulomb-type operators this is a `$t$-transformation' (for which we are not aware of any common name), and for which the variable in the inverse space was labelled with $\xi=t$. 
Expectation values that contain the cusp effects are obtained by computing the integral as the sum of a short-, $0\leq \xi < \xi_\Lambda$, and a long-range, $\xi_\Lambda\leq \xi<\infty$, part. The short-range part is calculated by direct integration with the approximate wave function expressed with ECGs. 

We explained in this paper that the effect of the singular derivative of the wave function at the coalescence points, where the exact wave function cusps, appears in the asymptotic tail of the integrand in the inverse space. 
Using this connection and the analytic cusp conditions, we derived the analytic form of the long-range tail of the integrands  for the $p^2_i$, $1/r_{ij}$, $1/r_{iA}$-type operators and our derivation reproduced the asymptotic expressions for $p_i^4$, $\delta(r_{iA})$, and $\delta(r_{ij})$ of Ref.~\cite{pachucki_acceleration_2005}.
It is interesting to note that, in the inverse space, the asymptotic tail of the non-relativistic operators ($p^2$ and $1/r$) decay faster ($\sim 1/p^4$ and $\sim 1/t^3$), than the tail of the more
`singular' operators, $\hat{p}^4$ and $\hat{\delta}(\br)$,
$\sim 1/p^2$  and $\sim 1/t^2$, respectively.

Exploratory results were reported for the expectation values of the non-relativistic energy operators, for which, in principle, it should be possible to improve upon the non-relativistic energy with the inclusion of the cusp `effect'. The practical realization of this idea appears to be limited, for the moment, by particular details of the fitting procedure of the asymptotic tail.

{
We also use the IT technique in this work to compute perturbative relativistic corrections for the ground state of H$_3^+$ near its equilibrium structure. We observe error cancellation among the singular terms in the perturbative Dirac--Coulomb energy, but for the perturbative Dirac--Coulomb--Breit energy the IT technique results in a 15~n$\Eh$\ improvement over the direct result. These perturbative relativistic energies pinpointed with the IT technique can be used for a detailed comparison in relation with the variational relativistic result of Ref.~\cite{jeszenszki_all-order_2021} that will be reported in future work.}

Finally, we would like to mention that the $\hat{p}^4$- and $\delta(\br)$-type singular operators appear not only in the 
perturbative relativistic theory but also in lower-bound theory 
due to the presence of the $\hat{H}^2$ operator \cite{weinsteinModifiedRitzMethod1934,suzuki_stochastic_1998,robbieirelandLowerBoundsAtomic2021}. This fact contributes to the observation that the energy lower bounds typically converge slower to the exact energy \cite{suzuki_stochastic_1998,robbieirelandLowerBoundsAtomic2021}, than the energy upper bound. It would be interesting to use (generalize) the IT technique to the $\hat{H}^2$ expectation value and variance computations, which may speed up the convergence of the best energy lower bounds \cite{pollakLowerBoundsCoulombic2021} and that would open the route to the computation of rigorous theoretical error bars for numerically computed non-relativistic energies.

\section*{Acknowledgments}
\noindent The authors thank Gustavo Avila for discussions about the quadrature integration. Financial support of the European Research Council through a Starting Grant (No.~851421) is gratefully acknowledged. RTI thanks the Erasmus+ program for funding a traineeship at ELTE.

\bibliography{BasisSetExpansion,book,ExplicitGauss,He,F12,LowerBound,MathAndConstants,Relativistic,transcorrelated}

\appendix
\setcounter{equation}{0}
\renewcommand{\theequation}{A\arabic{equation}}
\setcounter{table}{0}
\renewcommand{\thetable}{A\arabic{table}}
\setcounter{figure}{0}
\renewcommand{\thefigure}{A\arabic{figure}}

\section{Necessary Gaussian integrals for the short-range part \label{sec:Gint}}% \label{sec:shortintr}}
The approximate wave function is written as a linear combination of antisymmetrized products of $\chi$ spin and $\Theta$ ECGs functions, 
\begin{align}
  \Psi=\sum_{I=1}^{\Nbas} c_I \hat{\mathcal{A}}\lbrace \chi_I\Theta_I \rbrace \; 
\end{align}
with the $\hat{\mathcal{A}}=(\Nperm)^{-\frac{1}{2}}\sum_{p=1}^{\Nperm} \varepsilon_p \hat{P}_p$ antisymmetrization operator over the $\Nperm$ possible permutations with $\varepsilon_p$ parity.
Expectation values of a (permutationally invariant) $\hat{O}$ operator can be calculated as 
\begin{align}
  \langle %
    \Psi | \hat{O} | \Psi 
  \rangle
  =
  \sum_{I=1}^{\Nbas} 
  \sum_{J=1}^{\Nbas}
  \sum_{p=1}^{\Nperm}
    c_I^\ast c_J \varepsilon_{IJp}
    \langle
      \Theta_I | \hat{O} | \Theta_{J_p}
    \rangle
\end{align}
where $\varepsilon_{IJp}$ contains the parity of the permutation and the spin integrals,
and we need to calculate matrix elements of $\hat{O}$ with the ECG functions $\Theta_I$ and $\Theta_{Jp}$. Particle permutation leaves the mathematical form of the ECG unchanged, and assumes transformation of the $\bos{A}$ and $\bos{s}$ parameter arrays (for further details, see for example, Ref.~\cite{matyus_molecular_2012}). 

%%%%%%%%%%%%%%%%%%%%%%%%%%%%%%%%%%%%%%%%%%%%%%%%%%%%%%%%%%
During the IT procedure, the short-range part of the expectation values is computed by direct integration with the basis functions. For the short-range calculations, the following integrals were used.

\subsection{Coulomb integral over the short-range interval}
Using the following notations:
\begin{align}
\bos{e}_i &= \underline{\bos{A}}_i \bos{s}_i \ ,\\
\bos{e}_{ij} &= \bos{e}_i + \bos{e}_j \ , \\
\eta_{ij} &= \bos{s}_i^T \underline{\bos{A}}_i \bos{s}_i + \bos{s}_j^T \underline{\bos{A}}_j \bos{s}_j  \ , \\
\bos{A}_{ij} &= \bos{A}_i + \bos{A}_j \ , \\
\gamma_{ij} &= \bos{e}_{ij}^T \underline{\bos{A}}_{ij}^{-1} \bos{e}_{ij}- \eta_{ij} \ , \\
%    \bos{J}_{12} &= \left ( \begin{array}{rr}
%        1 & -1 \\
%        -1 & 1
%    \end{array}\right) \ , \\
  (\bos{J}_{12})_{kl} 
  &= \delta_{1k}\delta_{1l} 
  +  \delta_{2k}\delta_{2l} 
  -  \delta_{1k}\delta_{2l} 
  -  \delta_{2k}\delta_{1l}
    , \quad k,l=1,\ldots,N  
    \\
\beta_{ij} &= \be_{ij}^T  \underline{\bA}_{ij}^{-1} \underline{\bJ}_{12} \underline{\bA}_{ij}^{-1} \be_{ij} \ , \\
a_{ij} &= \Tr \left( \bJ_{12} \bA_{ij}^{-1}  \right) \ , \\
S_{ij} &= \exp \left( \gamma_{ij} \right) \frac{\pi^\frac{3N}{2}}{|\bA_{ij}|^{3/2}} \ ,
\end{align}
the Coulomb integral for a finite range can be given explicitly as,
\begin{align}
  \matrixel**{\Theta_i}{\left(\frac{1}{r_{12}}\right)_\Lambda}{\Theta_j} &= \frac{2}{\sqrt{\pi}} \int_0^\Lambda \dd t\ \matrixel**{\Theta_i}{\eem^{-r_{12}^2 t^2} }{\Theta_j} \nonumber \\
  &=\frac{2}{\sqrt{\pi}}S_{ij} \int_0^\Lambda \dd t\ (1+t^2 a_{ij})^{-3/2} \eem^{-\frac{\beta_{ij} t^2}{1+t^2a_{ij}}} \nonumber \\
  &=\frac{S_{ij}}{\sqrt{\pi\beta_{ij}}} \int_0^{\frac{\Lambda^2\beta_{ij}}{1+\Lambda^2a_{ij}}} \dd z\ z^{-\frac{1}{2}}\eem^{-z}=\frac{S_{ij}}{\sqrt{\beta_{ij}}} \erf\left[\left( \frac{\Lambda^2\beta_{ij}}{1+\Lambda^2a_{ij}} \right)^\frac{1}{2}\right] \ .
\end{align}
We note that the $t$ dependence of the short-range $\delta(\br)$
was integrated by Gauss--Legendre quadrature.

\subsection{Momentum integrals}
In this subsection, we draft the integration of the angular degrees of freedom 
for the momentum density, which is used in the second step of Eq.~(\ref{pkexpval}),
\begin{align}
  \int \dd \bp_1\ p_1^k\ \bar{\rho}(\bp_1) 
  =
  4\pi \int_0^\infty \dd p\ p^{k+2}\ 
  %\bar{\rho}(p) \ .
  \langle \bar{\rho}\rangle_{\theta,\phi}(p) \ .
\end{align}
To work out this step, we write down the integral for two basis functions in Fourier space that is proportional with
(where $a\in\mathbb{R}^+$, $\bos{d}\in\mathbb{R}^3$, and $d=|\bos{d}|$ are constant coefficients containing the exponent matrices and shift vectors of the basis functions)
\begin{align}
  \int\dd\bp_1\ p^k_1\ 
    \exp\left(
      -a p_1^2 +\iim\boldsymbol{d}^\text{T}\bp_1
      \right)
  &=
  \int_0^{2\pi} \dd \phi \int_0^\infty \dd p\ p^{2+k}
  \int_0^\pi \dd\theta \sin \theta 
    \exp\left(%
      -a p^2 + \iim |\boldsymbol{d}| p \cos \theta \right) 
  \nonumber \\
  &=
  2\pi 
  \int_0^\infty \dd p\ 
    p^{k+2}
  \int_{-1}^1 \dd z\ 
    \exp\left(%
      -a p^2 + \iim d p z \right) 
  \nonumber \\
  &=
  \frac{4\pi}{d}  
  \int_0^\infty \dd p\ 
    p^{k+1}
    \sin(d p)\  
    \text{e}^{-a p^2} \; .
  \label{eq:pangint}
\end{align}
We note that $\iim\bos{d}$ is purely imaginary for any configuration-space shift vectors, $\bos{s}\in\mathbb{R}^{3N}$, due to Eq.~(\ref{eq:FTecg}).
The short-range part of the integral in Eq.~(\ref{eq:pangint}) can be calculated analytically which we show for $k=2$:
\begin{align}
    \langle p^2 \rangle_\Lambda&=\frac{4\pi}{d}\int_0^\Lambda \dd p\ p^3\eem^{-a p^2}\sin(dp) = \frac{4\pi}{d} \partial_a \partial_d \int_0^\Lambda \dd p\ \eem^{-ap^2}\cos(dp) \nonumber \\
    &=\frac{4\pi}{2d} \partial_a \partial_d \int_0^\Lambda \dd p\ \left[ \eem^{-ap^2+\iim dp} + \eem^{-ap^2-\iim dp} \right] \nonumber \\
    &=\frac{\pi^{3/2}}{d\sqrt{a}}\partial_a\partial_d\eem^{-\frac{d^2}{4a}} \left[ \erf\left(\sqrt{a}\Lambda+\frac{\iim d}{2\sqrt{a}}\right) + \erf\left(\sqrt{a}\Lambda-\frac{\iim d}{2\sqrt{a}}\right) \right] \nonumber \\
    &= \frac{1}{8a^{7/2}d}\left\lbrace d(6a-d^2)\pi^{\frac{3}{2}}\eem^{-\frac{d^2}{4a}} \left[\erf\left(\sqrt{a}\Lambda+\frac{\iim d}{2\sqrt{a}}\right) + \erf\left(\sqrt{a}\Lambda-\frac{\iim d}{2\sqrt{a}}\right) \right] \right. \nonumber \\
    &\quad-4\pi\sqrt{a}\eem^{-a\Lambda^2}\left[2ad\Lambda\cos(d\Lambda)+\left(4a+4a^2\Lambda^2-d^2\right)\sin(d\Lambda)\right]\left. \vphantom{\frac{1}{1}}\right\rbrace \; .
    \label{eq:ptwo}
\end{align}
If the ECGs are centered at the origin of the coordinate system, we need to consider the $d\rightarrow 0$ limit of the general expression:
\begin{align}
    \lim_{d\rightarrow 0} \langle p^2 \rangle_\Lambda&= \frac{3\pi^{3/2}}{2a^{5/2}} \erf\left(\sqrt{a}\Lambda\right)-\frac{\pi\Lambda}{a^2}\left(3+2a\Lambda^2\right)\eem^{-a\Lambda^2} \; .
    \label{eq:ptwos}
\end{align}

\section{Connection between wave function derivatives in real space and the decay rate of
the asymptotic tail in momentum space \label{sec:smoothness}}
%Let's first consider a one-dimensional general function and later generalize the observations to $| \Psi \rangle $ in thee dimensions with Coulomb interactions.
%
Let us consider an $L^2$ integrable function, $f(x)$, which decays to zero for $x\rightarrow \pm\infty$. Moreover, its $(k-1)$th derivative is discontinuous at $x_0$, and its $k$th
derivative at this point is related to the Dirac delta function
\footnote{A discontinuous function cannot be
differentiated in a rigorous way. However, the differentiation can be generalized using the so-called weak derivative \cite{ranadeFunctionalAnalysisQuantum2015}, which can be calculated for these functions. This leads to the expected Dirac delta function as the weak derivative of the Heaviside step function.}
\begin{align}
  \frac{\mbox{d}^k f(x)}{\mbox{d} x^k} 
  \sim 
  \delta(x-x_0)A(x)  \label{pderddelta} \ ,
\end{align}
where $A(x)$ is a continuous regular function, which describes the
$k$th derivative everywhere else. 
Next, let us consider the Fourier transform of $f(x)$ and its momentum-space properties,
\begin{align}
  \tilde{f}(p) 
  =  
  \frac{1}{ \sqrt{2 \pi}}\int\limits_{-\infty}^{\infty} \mbox{d}x \, f(x) \eem^{-\iim p x} \ .
\end{align}
Using partial integration, $\bar{f}(p) $ can be expressed with the integral
of the derivative of $f(x)$,
\begin{align}
\tilde{f}(p)= 
-\frac{1}{ \sqrt{2 \pi} \iim p} \underbrace{\left[ f(x) \eem^{-\iim p x} \right]_{-\infty}^{\infty}}_{0}
+\frac{1}{ \sqrt{2 \pi} \iim p} \int\limits_{-\infty}^{\infty} \mbox{d}x \, \frac{\mbox{d} f(x)}{\mbox{d} x} \eem^{-\iim p x} \ ,
\label{eq:partint1Done}
\end{align}
where the first term in the right hand is zero, since our original condition was $\lim\limits_{x\rightarrow \pm \infty }f(x)=0$. The partial integration can be repeated $k$ times,
\begin{align}
\tilde{f}(p) 
=  
\frac{1}{ \sqrt{2 \pi} }\int\limits_{-\infty}^{\infty} 
  \mbox{d}x \, f(x) \eem^{-\iim p x} =
\frac{1}{\sqrt{2 \pi}}\left( \frac{-\iim}{p }\right)^k \int\limits_{-\infty}^{\infty} \mbox{d}x \, 
\frac{\mbox{d}^k f(x)}{\mbox{d} x^k} \eem^{-\iim p x}
=  \frac{1}{ \sqrt{2 \pi}}\left( \frac{-\iim}{p}\right)^k \eem^{-\iim p x_0} A(x_0) \ ,
\label{eq:partint1Dmulti}
\end{align}
where \rref{pderddelta} is used and we assumed that the Dirac delta
predominantly determines the integral expression above.  Since
$\eem^{-\iim p x_0}$ is bounded,
\begin{align}
\left| \eem^{-\iim p x_0} \right| = 1 \ ,
\end{align}
in the limit of large $p$ values, $\bar{f}(p)$ decays polynomially,
\begin{eqnarray}
  p>p_\Lambda: \quad
  \bar{f}(p) 
  %\stackrel{p \rightarrow \infty }{\sim} 
  \sim
  \frac{1}{p^k} \ . \label{relbetweencjandkj}
\end{eqnarray}
\vspace{0.5cm}

\section{Fourier transformation of \texorpdfstring{$r f( \vartheta , \varphi )$}{} \label{sec:FTrf} }

In this Appendix, we consider the effect of the function $f(\vartheta,\varphi)$ in \rref{cuspforwf} on the integrand values. The function $f(\vartheta,\varphi)$ can be written as a linear combination of $Y_{1m}$ spherical symmetric functions. In what follows we show that the Fourier transform 
of $r f(\vartheta,\varphi)$ is local, moreover, its contribution is zero in the asymptotic tail of the kinetic and mass-velocity term integrands.
So, we consider %it is sufficient to consider $r Y_{1m}(\vartheta,\varphi)$,
\begin{align}
    h(\bos{p})=\frac{1}{\sqrt{8 \pi^3}}
    \int \dd \bos{r} \, 
      \eem^{\iim \bos{p} \bos{r}} r Y_{1m}(\vartheta,\varphi) \ . \label{FTrY}
\end{align}
In order to perform the Fourier transformation let us expand the plane wave in terms of spherical harmonics \cite{f.w.j.olverNISTDigitalLibrary2021}, 
\begin{align}
    e^{\iim \bos{p} \bos{r}} = 4 \pi\sum_{\ell=0}^\infty \sum_{m=-\ell}^\ell \iim^\ell j_\ell(pr) Y_{lm}^*\left( \frac{\bos{p}}{p} \right)  Y_{lm}\left( \frac{\bos{r}}{r} \right) \ , \label{SHexp}
\end{align}
where $j_\ell(x)$ is the spherical Bessel function \cite{e.w.weissteinSphericalBesselFunction2021}. Substituting \rref{SHexp} into \rref{FTrY} and using the orthogonality relation between the spherical harmonics, the angular integral can be evaluated, and we obtain
\begin{align}
    h(\bos{p})=\frac{ \iim  }{ \sqrt{2 \pi} }  Y_{lm}^*\left( \frac{\bos{p}}{p} \right)      \int \dd r \,   r^3 j_1\left( pr \right) \ . \label{FTrad}
\end{align}
Using the identity, 
\begin{align}
    \frac{\partial}{\partial p}  j_0(pr)= - r j_1\left( pr \right) \ , 
\end{align}
which can be checked by substituting the explicit expressions for the spherical Bessel functions \cite{e.w.weissteinSphericalBesselFunction2021}. We can rewrite the integral in \rref{FTrad} as
\begin{align}
    h(\bos{p})=- \frac{ \iim  }{ \sqrt{2 \pi}} Y_{lm}^*\left( \frac{\bos{p}}{p} \right) \frac{\partial}{\partial p} \int \dd r \,   r^2 j_0\left( pr \right) \ .
\end{align}
Then, we can recognize one of the identities  of the Dirac delta function \cite{uginciusIntegralRepresentationDirac1972}, $ \delta(p)=\frac{2 p^2}{\pi} \int \dd r \,   r^2 j_0\left( pr \right) $,
\begin{align}
    h(\bos{p})=-  \frac{\sqrt{2} \iim  }{\pi^{3/2}}Y_{lm}^*\left( \frac{\bos{p}}{p} \right) \frac{\partial}{\partial p} \frac{\delta(p)}{p^2} \ ,
\end{align}
and the differentiation can be performed by using the identity $\delta(x)=-x \delta'(x)$ for the derivative of the Dirac delta,
\begin{align}
    h(\bos{p})=\frac{3 \sqrt{2} \iim}{\pi^{3/2}}  \frac{\delta(p)}{p^3}  Y_{lm}^*\left( \frac{\bos{p}}{p} \right)  \ . \label{FTrfFinal}
\end{align}
The appearance of $\delta(p)$ ensures that the resulting function is localized near the origin (small $p$ values), and thus, it does not contribute to the large-$p$ asymptotic tail.

\section{Determination of the \texorpdfstring{$\rho(\bos{0})$}{} and \texorpdfstring{$\eta(\bos{0})$}{} values for the ground state of the He atom\label{sec:rhoeta}}

\begin{figure}
   \includegraphics[scale=1]{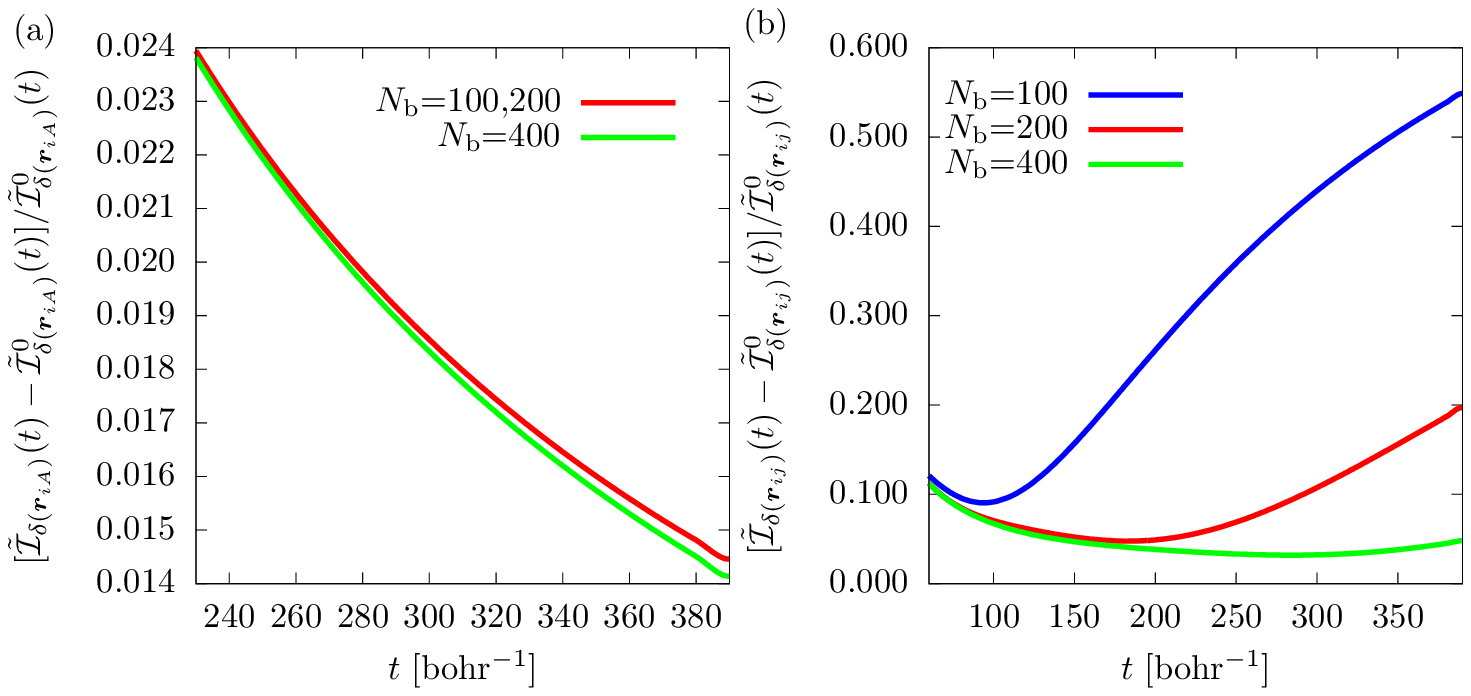}
  \caption{%
    Relative difference in the asymptotic tail of 
    the numerically calculated functions, $\itfun{\delta(\bos{r}_{iA})}(t)$ and  $\itfun{\delta (\bos{r}_{ij})}(t)$, 
    and 
    the analytic-leading order expressions of the asymptotic tail, $\titfun{\delta(\bos{r}_{iA})}^0(t)=8 \rho(\bos{0})/(\sqrt{\pi}t^2)$ and $\titfun{\delta(\bos{r}_{ij})}^0(t)=-2 \eta(\bos{0})/(\sqrt{\pi}t^2)$
    for the example of the ground state of the helium atom with an increasing number of ECG basis functions ($N_\text{b}$). 
  \label{fig:asymdelta}
  }
\end{figure}

In order to determine accurate values for $\delta(\br_{iA})=N^{-1}\rho(\bos{0})$ and $\delta(\br_{ij})=[N(N-1)]^{-1}\eta(\bos{0})$, the expectation values of $\delta(\br_{iA})$ and $\delta(\br_{ij})$ are obtained in an iterative procedure (Sec.~\ref{sec:ITCD}). 
The grid points used in the fitting are selected according to Sec.~\ref{sec:nonrel}. 
For $\delta(\br_{iA})$ and $\delta(\br_{ij})$, the fitting intervals start at 5 bohr$^{-1}$, and at 1 bohr$^{-1}$, respectively, which are sufficient to avoid complicated correlation effects at low $t$ values (see also Fig.~\ref{fig:itcurvesHe}).    
For the asymptotic range, the relative deviation of the integrands from the leading-order analytic terms is shown in Figure~\ref{fig:asymdelta}. The function $\itfun{\delta(\bos{r}_{iA})}$ appears to be robust with respect to the number of basis functions, while 
$\itfun{\delta(\bos{r}_{ij})}$ is more sensitive to the basis set.

After inspection of these figures, we set $t_\Lambda=100$ bohr$^{-1}$ for the upper end of the interval used for the fitting, and the beginning of the long-range integral.
The $\rho(\bos{0})$ and $\eta(\bos{0})$ values obtained in this computational setup with seven fitting parameters are collected in Table~\ref{tab:rhoeta}.

\begin{figure}
  \includegraphics[scale=1]{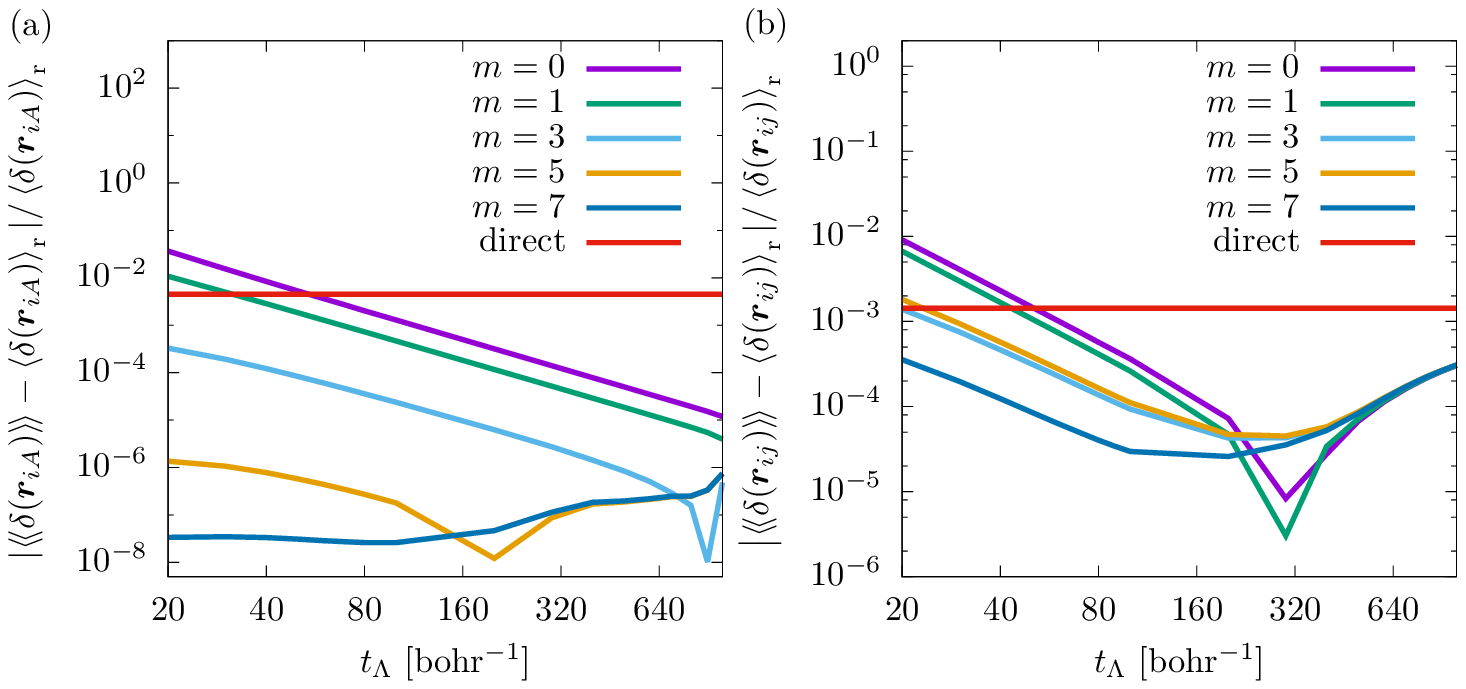}
  \caption{%
Relative error of $\llangle\delta(\bos{r}_{iA})\rrangle$ and $\llangle\delta(\bos{r}_{ij})\rrangle$ for various $t_\Lambda$ threshold values and $m$ terms in the fitting, \rref{eq:asymdelta}, using 400 ECGs. The result `direct'
was obtained by numerical integration over the entire $t$ range. 
The reference values are
$\langle \delta(\bos{r}_{iA}) \rangle_\mathrm{r}=3.620\ 858\ 637\ 7(3)$~bohr$^{-1}$
and
$\langle \delta(\bos{r}_{ij}) \rangle_\mathrm{r}=0.106\ 345\ 370\ 636(2)$~bohr$^{-1}$
 \cite{pachucki_acceleration_2005}. 
  \label{fig:deltaerr}
  }
\end{figure}

Figure~\ref{fig:deltaerr} shows the relative error of 
$\llangle \delta(\br_{iA})\rrangle$ 
and 
$\llangle \delta(\br_{ij})\rrangle$ 
in comparison with data available from Ref.~\cite{pachucki_acceleration_2005}.

\begin{table}
  \caption{%
    Convergence of the density and the pair correlation functions, in bohr$^{-3}$,  at the coalescence point for the ground-electronic state of the helium atom computed with the IT technique.  $\nb$ is the number of the basis functions. $t_\Lambda=100$ bohr$^{-1}$. 
    \label{tab:rhoeta}
  }
  \begin{tabular}{@{}l l l  @{}}
    \hline\hline\\[-0.35cm] 
    \multicolumn{1}{l}{$\nb$} &
     \multicolumn{1}{c}{$\rho(\bos{0})$} &
    \multicolumn{1}{c}{$\eta(\bos{0})$ }  \\
    \cline{1-3}\\[-0.3cm]
     100 &  3.620 845 647    &  0.106 366 877   \\
     200 &  3.620 857 171    &  0.106 350 118  \\
     400 &  3.620 858 545    &  0.106 348 521 \\ \hline
     %{\color{red}Ref.~\cite{pachucki_acceleration_2005} &  3.620 858 637 7(3)   & 0.106 345 370 636 (2)} \\
     Ref.~\cite{drake_high_2006} & 3.620 858 636 98(6)    & 0.106 345 371 2(2) \\
    \hline\hline
  \end{tabular}
\end{table}

\end{document}